\documentstyle[bbox,epsf,cite]{AnnRev}
\begin{document}
\def\slash#1{\rlap{$#1$}/} 
\def\gapp{\, \raisebox{-.5ex}{$\stackrel{>}{\scriptstyle\sim}$}\, }
\def\lapp{\, \raisebox{-.5ex}{$\stackrel{<}{\scriptstyle\sim}$} \, }
\def\half{\textstyle{1\over 2}}

\pagestyle{myheadings} 
\markboth{\rm ULRICH HEINZ \& BARBARA V. JACAK}{\rm TWO-PARTICLE
  CORRELATIONS IN HEAVY-ION COLLISIONS}
\title{TWO-PARTICLE CORRELATIONS IN RELATIVISTIC HEAVY-ION COLLISIONS}
\author{Ulrich Heinz
\affiliation{Theory Division, CERN, CH-1211 Geneva 23, Switzerland,
  ulrich.heinz@cern.ch,\\
  and Institut f\"ur Theoretische Physik, Universit\"at Regensburg,
  D-93040 Regensburg, Germany} 
  Barbara V. Jacak
\affiliation{Department of Physics, SUNY at Stony Brook,
  Stony Brook, NY 11794, USA, jacak@skipper.physics.sunysb.edu}}
\begin{keywords}
Hanbury Brown - Twiss (HBT) interferometry, Bose-Einstein correlations,
two-particle correlations, three-particle correlations, quark-gluon 
plasma, collective expansion flow, source sizes and lifetimes,
freeze-out 
\end{keywords}
\begin{abstract}
Two-particle momentum correlations between pairs of identical
particles produced in relativistic heavy-ion reactions can be analyzed
to extract the space-time structure of the collision fireball. We
review recent progress in the application of this method, based on
newly developed theoretical tools and new high-quality data from
heavy-ion collision experiments. Implications for our understanding of
the collision dynamics and for the search for the quark-gluon plasma
are discussed.  
\end{abstract}
\maketitle

\section{INTRODUCTION AND OVERVIEW}
\label{sec1}

\subsection{Intensity Interferometry}
\label{sec1.1}

The method of two-particle intensity interferometry was discovered in
the early 1950's by Hanbury Brown and Twiss (HBT) \cite{HBT} who
applied it to the measurement of the angular diameter of stars and
other astronomical objects. These first measurements used two-photon
correlations. An independent development occurred in the field of
particle physics in 1960 by Goldhaber, Goldhaber, Lee and Pais
\cite{GGLP} who extracted from two-pion correlations the spatial 
extent of the annihilation fireball in proton-antiproton reactions. 
The method exploits the fact that identical particles which sit nearby
in phase-space experience quantum statistical effects resulting from
the (anti)symmetrization of the multiparticle wave function. For bosons,
therefore, the two-particle coincidence rate shows an enhancement at
small momentum difference between the particles. The momentum range of
this enhancement can be related to the size of the particle source in
coordinate space.  

HBT interferometry differs from ordinary amplitude interferometry that
it compares {\em intensities} rather than amplitudes at different
points. It shows the effects of Bose or Fermi statistics even
if the phase of the (light or matter) wave is disturbed by
uncontrollable random fluctuations (as is, for example, the case for
starlight propagating through the earth's atmosphere) or if the
counting rate is very low. To illustrate how the method works and how
its applications in astro\-no\-my and in particle physics differ let us
consider the following simple model \cite{B69,B98}: two random 
point sources $a$ and $b$ on a distant emitter, separated by the
distance $\bbox{R}$, emit identical particles with identical energies
$E_p=(m^2+p^2)^{1/2}$ which, after travelling a distance $L$, are
measured by two detectors 1 and 2, separated by the distance 
%
\begin{figure}[ht]
\centerline{\epsfxsize=13cm\epsffile{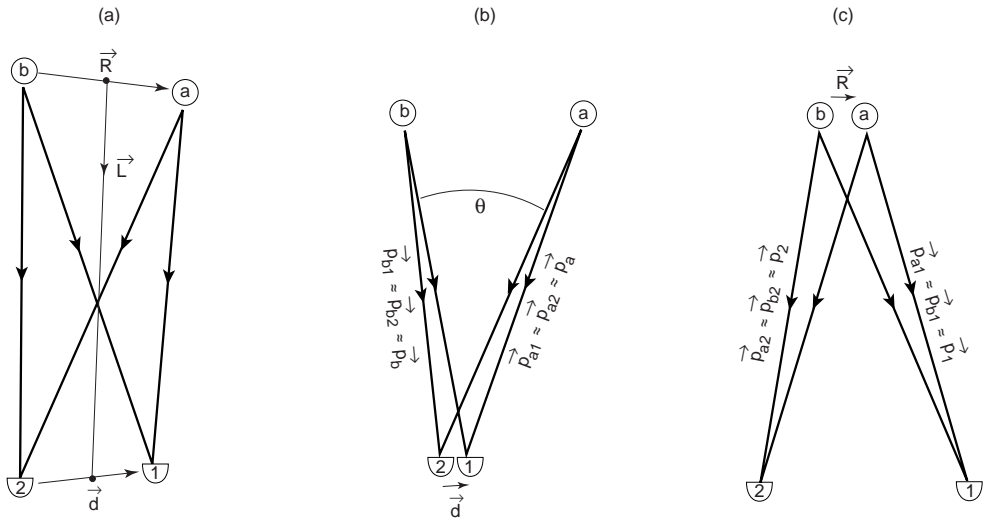}}
\caption{Measurement of the separation $\protect\bbox{R}$ of two
  sources $a$ and $b$ by correlating the intensities in detectors 1
  and 2 at varying distances $\protect\bbox{d}$. {\bf a}: The general
  scheme. {\bf b}: The specific situation in astronomy. {\bf c}: The
  specific situation in particle physics.     
\label{F0}}
\end{figure} 
%
$\bbox{d}$ (see Figure~\ref{F0}a). $L$ should to be much larger
than $R$ or $d$. The total amplitude measured at detector 1 is then
 \begin{equation}
 \label{B1}
   A_1 = {1\over L}\Bigl( \alpha\, e^{i(pr_{1a}+\phi_a)} 
            + \beta\, e^{i(pr_{1b}+\phi_b)} \Bigr),
 \end{equation}
where $\alpha,\beta$ are the amplitudes emitted by points $a$ and
$b$, $\phi_a,\phi_b$ are their random phases, $r_{1a},r_{1b}$ are
their distances to detector 1. The total intensity in 1 is 
 \begin{equation}
 \label{B2}
   I_1 = {1\over L^2}\Bigl( \vert\alpha\vert^2 + \vert\beta\vert^2 
         + 2\,{\rm Re\, } \alpha^*\beta\,
         e^{i[p(r_{1b}-r_{1a})+(\phi_b-\phi_a)]} \Bigr),
 \end{equation}
with a similar result for $I_2$. The last term in (\ref{B2}) contains
information on the distance $R$ between the sources $a$ and $b$, but
it vanishes after averaging the signal over some time, {\it i.e.} over
the random phases $\phi_{a,b}$: 
 \begin{equation}
 \label{B3}
   \langle I_1\rangle = \langle I_2\rangle = 
   {1\over L^2}\Bigl( \vert\alpha\vert^2 + \vert\beta\vert^2 \Bigr).
 \end{equation}
The product of the averaged intensities $\langle I_1\rangle
\langle I_2\rangle$ is thus independent of both $R$ and $d$.

The same is not true for the time-averaged {\em coincidence rate}
which is obtained by multiplying the two intensities before averaging:
 \begin{equation}
 \label{B4}
   \langle I_1 I_2\rangle = \langle I_1\rangle \langle I_2\rangle +
   {2\over L^4} \vert\alpha\vert^2 \vert\beta\vert^2 
   \cos\Bigl(p(r_{1a}-r_{2a}-r_{1b}+r_{2b})\Bigr).
 \end{equation}
The two-particle intensity correlation function is thus given by
 \begin{equation}
 \label{B5}
   C(\bbox{R},\bbox{d}) = 
   {\langle I_1 I_2\rangle \over \langle I_1\rangle \langle I_2\rangle}
   = 1 + {2 \vert\alpha\vert^2 \vert\beta\vert^2 \over
            (\vert\alpha\vert^2 + \vert\beta\vert^2)^2}
   \cos\Bigl(p(r_{1a}-r_{2a}-r_{1b}+r_{2b})\Bigr).
 \end{equation}
For large $L\gg R,d$, the argument of the second, oscillating term becomes
 \begin{equation}
 \label{B6}
   r_{1a}-r_{2a}-r_{1b}+r_{2b} \longrightarrow {d\,R\over L}
   \Bigl( \cos(\bbox{d},\bbox{R}) - \cos(\bbox{d},\bbox{L})
          \cos(\bbox{R},\bbox{L}) \Bigr).
 \end{equation}
Note the symmetry of this expression in $d$ and $R$, the separations of
the detectors and of the emitters; this symmetry is lost in the two 
practically relevant limits: 

\begin{enumerate}
 
\item
In astronomical applications the emission points $a$ and $b$ are part
of a star's surface or of even larger objects, while the detectors on
earth are only a few meters or kilometers apart: $R\gg d$. In this limit 
(see Figure~\ref{F0}b) the cosine-term in (\ref{B5}) reduces to
$\cos(\bbox{d}{\cdot}(\bbox{p}_a{-}\bbox{p}_b))$, with $\bbox{p}_{a,b} = p\, 
\bbox{e}_{a,b}$, the unit vectors $\bbox{e}_{a,b}$ giving the
directions from the detectors to the two emission points $a,b$. 
Experimentally one varies the distance $\bbox{d}$ between the
detectors, and from the resulting oscillation of the signal one
extracts the {\em angular} separation $\theta$ of the two emitters via 
$\vert \bbox{p}_a{-}\bbox{p}_b\vert \sim \theta/\lambda$. Absolute
determination of the source separation $R$ is possible only if their
distance $L$ is known from elsewhere.
 
\item
In nuclear and particle physics the emitters are very much smaller
than the detector separation, $R\ll d$. Then (see Figure~\ref{F0}c)
the cosine-term in (\ref{B5}) becomes 
$\cos(\bbox{R}{\cdot}(\bbox{p}_1{-}\bbox{p}_2))$, in similar notation
as above. Now the experimental control variable is the momentum
difference $\bbox{p}_1{-}\bbox{p}_2$, and $\bbox{R}$ is extracted from
the oscillation period. 

\end{enumerate}
In real life one has, instead of two discrete ones, a continuum
of sources described by a distribution $\rho(\bbox{R})$ of their
relative distances. For the case $R\ll d$, after averaging (\ref{B5})
over this relative distance distribution, the measured correlation
function is then given by the Fourier transform of $\rho(\bbox{R})$:
 \begin{equation}
 \label{B7}
   C(\bbox{p}_1-\bbox{p}_2) - 1 \sim 
   \int d^3R\, \rho(\bbox{R})\, 
   \cos(\bbox{R}\cdot(\bbox{p}_1-\bbox{p}_2))\, .
 \end{equation}
As we will see, this expression is only applicable to static sources.
The assumption of a static source is adequate for stars. The particles
emitted in high energy hadron or nuclear reactions, however, come from
sources which feature strong dynamical evolution both along the beam
direction and transverse to it. As a result, two-particle correlation
measurements in heavy-ion physics exhibit a much {\em richer}
structure, but their interpretation is also considerably more
involved. The present review covers the technical tools required
to access this richness of information and their practical application 
in heavy-ion reactions.


\subsection{HBT Interferometry for Heavy-Ion Collisions}
\label{sec1.2}

High energy heavy-ion collisions produce hadronic matter under 
intense conditions of temperature and density. While the highest 
densities are reached in the early stages of the collision, most 
of the observed particles are hadrons, which are emitted rather 
late in the evolution of the collision. For this reason, the 
measured momentum spectra and correlations contain direct information 
only about the size, shape, and dynamics of the source at ``freeze-out'', 
{\it i.e.} when the hadrons cease to interact. 

The dynamical information is of particular importance as it allows us
to connect the observed final state with the hot and dense early
stages. Therefore much of the effort in the last few years has gone
into the extraction of this dynamics. It turns out that both the
single-particle spectra and two-particle correlations are sensitive to
certain combinations of thermal and collective motion in the source. A
simultaneous analysis of particles with different masses allows for a
separation of the two effects: while the thermal motion generates a
common momentum distribution in the source, the collective motion
creates a flow velocity which combines with the thermal momentum
proportionally to the particle's mass. Further discrimination is
achieved by combining the spectra with the two-particle correlations
which reflect the collective source dynamics through a characteristic
momentum dependence of their width parameters.

The aim of HBT (or two-particle Bose-Einstein) interferometry is
therefore to provide, in conjunction with a simultaneous analysis of
single-particle spectra, a complete and quantitative characterization
(both geometrically and dynamically) of the emitting source at
``freeze-out''. Such a characterization can be used for backward 
extrapolations into the interesting dense earlier stages of the
collision, and it provides stringent constraints on dynamical models
aiming to describe the kinetic evolution of these stages.


\subsection{New Developments During the Last Decade}
\label{sec1.3}

The last decade has brought great strides in measurement, theory, and
interpretation of two-particle correlations. Dedicated experiments,
optimized for momentum resolution, allow measurement of correlations
to very small momentum difference. Much increased statistics of
particle pairs has opened the possibility of multidimensional 
correlation analysis. Correlation functions of identified kaons have
become available; as many fewer kaons arise from resonance decays,
they provide a more direct picture of the emitting source than the
more prevalent pions. Correlations of proton pairs, which will not be
discussed here, are also measured in heavy-ion collisions. The
experimental programs at the CERN SPS and Brookhaven AGS have allowed
comparison between small and large systems at different energies,
using S and Pb beams at CERN and lower energy Si and Au beams at the
AGS. The comparisons are greatly aided by the development of a
commonly accepted analysis formalism among the experiments. Furthermore, 
the pair momentum dependence of the correlations is now being used to
provide dynamical information about the space-time structure of the
particle source. 

Intensive modeling with event generators, combined with methods to
extract correlation functions from them, has been used to study
experimental effects such as acceptance and to verify the usefulness
of simple model parameterizations in theory. New parametrizations of
the two-particle correlation function have been developed which are
particularly well adapted for the sources created in high energy
collisions. Recently, the necessity of separating geometrical,
temporal and dynamical aspects of the correlations has been
recognized, and methods to do so have been developed. Intensity
interferometry has thus developed into a quantitative tool which at
the present stage yields important source parameters at the level of
20\% accuracy. 


\section{THEORETICAL TOOLS FOR ANALYZING TWO-PAR\-TI\-CLE CORRELATIONS}
\label{sec2}
\subsection{1- and 2-Particle Spectra and Their Relation to the
            Emitting Source}
\label{sec2.1} 

The 2-particle correlation function is defined as the ratio of the 
Lorentz-invariant 2-particle coincidence cross section and the product 
of the two single particle distributions:
 \begin{equation}
 \label{1}
   C(\bbox{p}_1,\bbox{p}_2) 
   = {E_1 E_2 dN/(d^3p_1 d^3p_2) \over
      (E_1 dN/d^3p_1) (E_2 dN/d^3p_2)}\, . 
 \end{equation}
The single- and two-particle cross sections are normalized to
the average number of particles per event $\langle N \rangle$ and the
average number of particles in pairs $\langle N(N-1) \rangle$, respectively.
Different normalizations of the correlation function are sometimes 
used in the literature \cite{GKW79,MV97}. Careful consideration of 
the normalization is required when analyzing multi-particle symmetrization 
effects \cite{P93,CGZ95,CZ97,W98,ZSH98}.


\subsubsection{Pure quantum statistical correlations}
\label{sec2.1.1}

The most direct connection between the measured two-particle
correlations in momentum space and the source distribution in
coordinate space can be established if the particles are emitted
independently (``chaotic source'') and propagate freely from source 
to detector. Several approaches to derive this connection \cite{S73}
are worked out in the literature, parametrizing the source as a
covariant superposition of classical currents
\cite{GKW79,P84,KG86,PGG90,APW93,CH94,H97} or using a superposition of
nonrelativistic wave packets \cite{CZ97,W98}. One finds the simple
relations (with the upper (lower) sign for bosons (fermions))
 \begin{eqnarray}
   E_p {dN \over d^3p} &=&  \int d^4x\, S(x,p) \, ,
 \label{spectrum}\\
  C(\bbox{q},\bbox{K}) &=& 1 \pm 
  {\left\vert \int d^4x\, S(x,K)\, e^{iq{\cdot}x} \right\vert^2
   \over
   \int d^4x\, S(x,K+\half q) \ \int d^4y\, S(y,K-\half q)}\, ,
 \label{correlator}
 \end{eqnarray}
where the {\em emission function} $S(x,K)$ is an effective
single-particle {\em Wigner phase-space density} of the particles in
the source. (Wigner densities are real but not necessarily everywhere
positive.) For the single-particle spectrum (\ref{spectrum}) this
Wigner function must be evaluated on-shell, {\it i.e.} at 
$p^0{=}E_p{=}(m^2{+}\bbox{p}^2)^{1/2}$. The correlation function
(\ref{correlator}) was expressed in terms of the relative momentum 
$\bbox{q}{=}\bbox{p}_1{-}\bbox{p}_2$, $q^0{=}E_1{-}E_2$, and 
average (pair) momentum $\bbox{K}{=}(\bbox{p}_1{+}\bbox{p}_2)/2$, 
$K^0{=}(E_1{+}E_2)/2$. As the two measured particles are on-shell, 
$p^0_{1,2}{=}E_{1,2}{=}(m^2{+}\bbox{p}_{1,2}^2)^{1/2}$, the 4-momenta
$q$ and $K$ are {\em off-shell}. They satisfy the orthogonality relation   
 \begin{equation}
 \label{ortho}
   q \cdot K = 0\,.
 \end{equation}
That on the rhs of Eq.~(\ref{correlator}) one needs the emission 
function for off-shell momenta $K$ may at first seem troublesome.
In practice, however, the {\it on-shell appro\-xi\-mation} 
$K^0 \approx E_K{=}(m^2{+}\bbox{K}^2)^{1/2}$ is very accurate in 
heavy-ion collisions: the corrections of order $\bbox{q}^2/(8E_K^2)$ are 
small in the interesting domain of small relative momenta $\bbox{q}$, 
as a result of the large source sizes and the rest masses of the 
measured particles. A further simplification is achieved by making in 
the denominator of (\ref{correlator}) the {\em smoothness approximation} 
\cite{P95,P97}, taking the product of single-particle spectra at their 
average momentum $\bbox{K}$:
 \begin{equation}
  C(\bbox{q},\bbox{K}) \approx 1 \pm 
  \left\vert {\int d^4x\, S(x,K)\, e^{iq{\cdot}x} 
   \over
   \int d^4x\, S(x,K)} \right\vert^2  \equiv 1 \pm 
   \left\vert \langle e^{iq{\cdot}x} \rangle(K)\right\vert^2\, .
 \label{corrapp}
 \end{equation}
It is exact for exponential single-particle spectra, with corrections
proportional to their curvature in logarithmic representation.
Both approximations can {\em a posteriori} be corrected for in the
correlation radii (``HBT radii'', see below), using information from
the measured single-particle spectra \cite{CSH95a}. For heavy-ion
collisions, such corrections are usually negligible \cite{P97,CSH95a}.

We call $S(x,K)$ an {\em effective} single-particle Wigner density 
since different de\-ri\-va\-tions of the relation (\ref{correlator}) yield 
different microscopic interpretations for $S$. For a detailed 
discussion of this point we refer to \cite{WH99}. The differences 
can become conceptually important in sources with high phase-space 
density \cite{ZSH98}. So far, in heavy-ion collisions the phase-space 
densities at freeze-out appear to be low enough to neglect them 
\cite{M96,FTH98}.   
 

\subsubsection{The invertibility problem}
\label{sec2.1.2}

The mass-shell constraint (\ref{ortho}) eliminates one of the four
components of the relative momentum $q$; for example, it can be
resolved as
 \begin{equation}
   q^0 = \bbox{\beta}\cdot\bbox{q}\, , \qquad
   \bbox{\beta} = \bbox{K}/K^0 \approx \bbox{K}/E_K\, ,
 \label{mass-shell}
 \end{equation}
which gives the energy difference $q^0$ in terms of $\bbox{q}$ and the
velocity $\bbox{\beta}$ of the pair. With only three independent
$\bbox{q}$-components, the Fourier transform in (\ref{corrapp}) cannot
be inverted, {\it i.e.} the space-time structure of $S(x,K)$ cannot be fully
recovered from the measured correlator:
 \begin{equation}
   C(\bbox{q},\bbox{K}){-}1 \approx \pm
   \left\vert {\int_x e^{i\bbox{q}{\cdot}(\bbox{x}-\bbox{\beta}t)} S(x,K) 
   \over
   \int_x S(x,K)} \right\vert^2  = \pm  
   \left\vert {\int_x e^{i\bbox{q}{\cdot}\bbox{x}}
              S(t,\bbox{x}+\bbox{\beta}t;K) 
   \over
   \int_x S(x,K)} \right\vert^2 .
 \label{invert}
 \end{equation}
Separation of the spatial and temporal structure of the source thus
requires {\em additional model assumptions} about $S(x,K)$. 

We can connect (\ref{invert}) with (\ref{B7}) by introducing the 
{\em normalized relative distance distribution} which is a folding of 
the single-particle emission function with itself:
 \begin{equation}
 \label{reldis}
   d(x,K) = \int_X s\left(X+{\textstyle{x\over 2}},K\right)
                   s\left(X-{\textstyle{x\over 2}},K\right),\quad
   s(x',K) = {S(x',K)\over \int_{x'} S(x',K)}\, .
 \end{equation}
$d$ is an even function of $x$. It allows to rewrite \cite{WH99} 
the correlator in the form (\ref{B7}):
 \begin{equation}
 \label{reldis2}
   C(\bbox{q},\bbox{K})-1 \approx \pm \int d^4x\, \cos(q\cdot x)\,
   d(x,K) = \pm \int d^3x\, \cos(\bbox{q}\cdot\bbox{x}) \, 
   S_{\bbox{K}}(\bbox{x})\, .
 \end{equation} 
In the second equation we used (\ref{mass-shell}) and introduced
the {\em relative source function}
 \begin{equation}
 \label{relsource}
   S_{\bbox{K}}(\bbox{x}) = 
   \int dt\, d(t,\bbox{x}+\bbox{\beta}t;{\bbox{K}},E_K)\, .
 \end{equation}
In the pair rest frame where $\bbox{\beta}=0$, $S_{\bbox{K}}(\bbox{x})$ 
is the time integral of the relative distance distribution $d$, and
the time structure of the source is completely integrated out. On the
other hand, $S_{\bbox{K}}(\bbox{x})$ is, for each pair momentum
${\bbox{K}}$, fully reconstructible from the measured correlator
$C(\bbox{q},\bbox{K})$ by inverting the last Fourier transform in
(\ref{reldis2}). This ``imaging method'' was recently exploited in
\cite{BD97}. As we will see, interesting information about the source
dynamics and time structure is then hidden in the $\bbox{K}$-dependence 
of $S_{\bbox{K}}(\bbox{r})$; the latter can, however, not be 
unfolded without additional model assumptions about the source. 


\subsubsection{Final state interactions and unlike particle
  correlations} 
\label{sec2.1.3}

HBT measurements in high energy physics are mostly performed with
charged particles. These suffer long-range Coulomb repulsion
effects on the way from the source to the detector which, even for
boson pairs, cause a suppression of the measured correlator at
$\bbox{q}=0$. Moreover, the charged particle pair feels the total
electric charge of the source from which it is emitted. Final state
effects from strong interactions play a dominant role in proton-proton
correlations \cite{K77}, due to the existence of a strong $s$-wave
resonance in the two-nucleon channel just above threshold. Such final
state interaction (FSI) effects are sensitive to the average
separation of the two particles at their emission points and thus also
contain relevant information about the source size \cite{K77,BBM96,B98}. 

This has recently led to an increased effort to understand and exploit
FSI-induced two-particle correlations which also exist between pairs
of {\em unlike particles} \cite{LL82,AL96,VLPX97,Soff97,M98}. The
particular interest in such correlations arises from the fact that,
for particles with unequal masses, they allow under certain
circumstances to determine the {\em sign} of the time difference
between the emission points of the two particles or the {\em direction} 
of their separation at emission \cite{LL82,AL96,VLPX97,Soff97,M98};
this is not possible with correlations between identical particles. 
In most practical cases, however, the FSI-induced correlations are
considerably weaker than the Bose-Einstein correlations between pairs 
of identical particles. 

At the level of accuracy of Eqs.~(\ref{corrapp},\ref{reldis2}) which
use the smoothness approximation, the correlator can be easily
corrected for 2-body final state interactions by repla\-cing 
$e^{iq\cdot x}$ with a suitable distorted wave. Instead of 
(\ref{reldis2}) one thus obtains
\cite{K77} 
 \begin{equation}
 \label{Koonin}
   C(\bbox{q},\bbox{K}) = \int d^3r\, S_{\bbox{K}}(\bbox{r})\, 
   \left\vert \Phi_{\bbox{q}/2}(\bbox{r}) \right\vert^2\, .
 \end{equation} 
A slightly more general result which avoids the smoothness
approximation was derived in \cite{AHR98}. For simplicity the integral 
in (\ref{Koonin}) is written in coordinates of the pair rest frame
where $\bbox{K}=0$. $\Phi_{\bbox{q}/2}(\bbox{r})$ is an FSI distorted
scattering wave for the relative motion of the two particles with
asymptotic relative momentum $\bbox{q}$; for Coulomb FSI it is given
by a confluent hypergeometric function: 
 \begin{eqnarray}
   &&\Phi^{\rm Coul}_{\bbox{q}/2}(\bbox{r}) = \Gamma(1+\eta)\,
   e^{-{1\over 2}\pi\eta}\, e^{{i\over 2}\bbox{q}\cdot\bbox{r}}\,
   F(-i\eta;1;z_-)\, ,
 \label{Coulomb}\\
   &&z_\pm = {\textstyle{1\over2}}(qr\pm\bbox{q}\cdot\bbox{r}) 
   = {\textstyle{1\over2}}qr(1\pm\cos\theta) , \quad
   \eta = {\alpha\,m\over q} , 
 \label{Sommerfeld}
 \end{eqnarray}
where $\alpha=e^2/4\pi$. It describes the propagation of the particle
pair from an initial separation $\bbox{r}$ in the pair rest frame,
at the time when the second particle was emitted \cite{AHR98}, under 
the influence of the mutual FSI. For identical particle pairs it must
be properly symmetrized: $\Phi_{\bbox{q}/2}\mapsto 
{\textstyle{1\over\sqrt{2}}}(\Phi_{\bbox{q}/2}\pm\Phi_{-\bbox{q}/2})$. 
For a pointlike source $S_{\bbox{K}}(\bbox{x}) = \delta(\bbox{x})$
the correlator (\ref{Koonin}) with (\ref{Coulomb}) reduces to the
Gamov factor $G(\eta)$ (to $2G(\eta)$ for identical particles):
 \begin{equation}
 \label{Gamov}
   G(\eta) = \left\vert \Gamma(1+i\eta) e^{-{1\over 2}\pi\eta}
             \right\vert^2
           = {2\pi\eta\over e^{2\pi\eta} -1}\, .
 \end{equation}

For Coulomb FSI it was recently shown \cite{SLAPE98} that a very 
good approximation for the Coulomb correction can be taken from 
{\em measured} unlike-sign particle pairs in the following form:
 \begin{equation}
 \label{Sinyukov}
   C^{\pm \pm}_{\rm corr.}(\bbox{q},\bbox{K}) =
   {C^{+ -}_{\rm meas.}(\bbox{q},\bbox{K})\,
    C^{\pm \pm}_{\rm meas}(\bbox{q},\bbox{K}) \over
    G(\eta)\, G(-\eta)} \, .
 \end{equation}
The denominator (which deviates from unity for small $q \lapp 8m\alpha$)
corrects for the fact that even for a pointlike source the like-sign
and unlike-sign Coulomb correlations are not exactly each other's
inverse. The important observation in \cite{SLAPE98} is that this
correction is essentially independent of the source size.

The effects of the central charge of the remaining fireball on the
charged particle pair were studied in \cite{BBM96,B96}. For a static
source it was found that at large pair momenta the FSI reduces
(increases) the apparent size (HBT radius) for positively (negatively)
charged pairs \cite{BBM96,B96} whereas for small pair momenta the
apparent radius increases for both charge states \cite{B96}. Expanding
sources have not yet been studied in this context, nor has this effect
been quantitatively confirmed by experiment. Also, combining the
central interaction with two-body FSI remains an unsolved theoretical
challenge. 


\subsection{Source Sizes and Particle Emission Times from HBT 
            Correlations}
\label{sec2.2}

The two-particle correlation function is usually parametrized by a
Gaussian in the relative momentum components. We now discuss different
Gaussian parametrizations and establish the relationship of the
corresponding width parameters (HBT radii) with the space-time
structure of the source. 

\subsubsection{HBT radii as homogeneity lengths}
\label{sec2.2.1}

The space-time interpretation of the HBT radii is based on a Gaussian
approximation to the space-time dependence of the emission function
$S(x,K)$ \cite{CSH95b,CSH95a,CNH95,AS95,CL95,WSH96,HTWW96}. 
Characterizing the effective source of particles of momentum $K$ by its 
space-time variances (``rms widths'') 
 \begin{equation}
 \label{3.2}
   \langle \tilde x_\mu \tilde x_\nu \rangle (\bbox{K})
   \equiv \langle (x-\bar x)_\mu (x-\bar x)_\nu \rangle,
 \end{equation}   
where $\langle \dots \rangle$ denotes the ($K$-dependent) space-time
average over the emission function defined in (\ref{corrapp}) and
$\bar x(K) = \langle x \rangle$ is the center of the effective
source, one obtains from (\ref{corrapp}) the following generic
Gaussian form for the correlator \cite{CSH95b,CNH95,HTWW96}:
 \begin{equation}
    C(\bbox{q},\bbox{K}) = 1 \pm \exp\left[ - q_\mu q_\nu 
    \langle \tilde x^\mu \tilde x^\nu \rangle (\bbox{K}) \right]\, .
  \label{3.4} 
 \end{equation}
This involves the smoothness and on-shell approximations 
discussed in section \ref{sec2.1.1} which permit to write the
space-time variances $\langle \tilde{x}_{\mu} \tilde{x}_{\nu}\rangle$
as functions of $\bbox{K}$ only. Eq.~(\ref{3.4}) expresses the width 
of the correlation function in terms of the rms widths of the
single-particle Wigner density $S(x,K)$. Note that the absolute 
space-time position $\bar{x}(\bbox{K})$ of the source center does 
not appear explicitly and thus cannot be measured.

Instead of the widths of the single-particle function $S(x,K)$ we can 
also use the widths of the relative distance distribution $d(x,K)$ 
(see (\ref{reldis})) to characterize the correlation function. Starting
from (\ref{reldis}) one finds within the same Gaussian approximation
$C(\bbox{q},\bbox{K}) = 1 \pm \exp\left[-{1\over 2} q_\mu q_\nu
  \langle x^\mu x^\nu \rangle_d (\bbox{K}) \right]$; 
here $\langle\dots\rangle_d$ denotes the average with the relative 
distance distribution $d$. Since $d$ is 
even, $\langle x^\mu \rangle_d \equiv 0$. One sees that the rms widths 
of $S$ and $d$ are related by a factor 2: 
$\langle x^\mu x^\nu\rangle_d = 2 \langle \tilde x^\mu \tilde x^\nu\rangle_S$.
This shows that for a Gaussian parametrization of the correlator according 
to (\ref{3.4}), {\em without a factor ${1\over 2}$ in the exponent}, the 
width parameters are directly related to the rms widths of the 
single-particle emission function $S$ whereas a similar parametrization 
{\em which includes a factor ${1\over 2}$ in the exponent} gives as 
width parameters the rms widths of the relative distance distribution $d$ 
(or of the relative source function $S_{\bbox{K}}(\bbox{r})$). While the 
latter interpretation may be mathematically more accurate, the former is 
more intuitive and has therefore been preferred in the recent literature.  

In either case, the two-particle correlator yields
{\em rms widths of the effective source of particles with momentum
  $\bbox{K}$}. In general, these width parameters do not characterize
the total extension of the collision region. They rather measure the
size of the system through a filter of wavelength $\bbox{K}$. In the
language introduced by Sinyukov~\cite{S95} this size is the ``region
of homogeneity'', the region from which particle pairs with momentum
$\bbox{K}$ are most likely emitted. The space-time variances $\langle
\tilde{x}_{\mu} \tilde{x}_{\nu}\rangle$ coincide with total source
extensions only in the special case that the emission function shows
no position-momentum correlations and factorizes, $S(x,K) = f(x)\, g(K)$.

 
\subsubsection{Gaussian parametrizations and interpretation of HBT
               radii}
\label{sec2.2.2}

Relating (\ref{3.4}) to experimental data requires first the
elimination of one of the four $q$-components via the mass-shell
constraint (\ref{ortho}). Depending on the choice of the three
independent components different Gaussian parametrizations exist. 
 
A convenient choice of coordinate axes for heavy-ion collisions is the 
{\it osl}-system \cite{P84,BGT88}, with $l$ denoting the {\it longitudinal} 
(or $x_l$) direction along the beam, $o$ the {\it outward} (or $x_o$) 
direction along the transverse pair momentum vector $\bbox{K}_\perp$, 
and $s$ the third Cartesian direction, the {\it sideward} (or $x_s$) 
direction. In this system the {\it sideward} component $\beta_s$ of the 
pair velocity $\bbox{\beta}$ in (\ref{mass-shell}) vanishes by definition.   

The {\bf Cartesian parametrization}~\cite{P83} of the correlator (often 
referred to, historically somewhat incorrectly, as 
Pratt\cite{P84} - Bertsch\cite{BGT88} parametrization) is based on 
an elimination of $q^0$ in (\ref{3.4}) via (\ref{mass-shell}):
 \begin{equation}
 \label{Cartesian}
   C(\bbox{q},\bbox{K}) = 1 \pm \exp\left( - \sum_{i,j=o,s,l}
   R_{ij}^2(\bbox{K})\, q_i\, q_j\right).
 \end{equation}
The Gaussian width parameters (HBT correlation radii) $R_{ij}$ of the
Cartesian parametrization are related to the space-time variances of
the emission function by \cite{BDH94,CSH95b,HB95} 
 \begin{equation}
 \label{CartHBT}
   R_{ij}^2(\bbox{K}) = \left\langle (\tilde x_i - \beta_i \tilde t)
   (\tilde x_j - \beta_j \tilde t)\right\rangle, \quad
   i,j=o,s,l.
 \end{equation}
These are {\em 6 functions of three variables}: the pair rapidity $Y$, 
the modulus $K_\perp$ and the azimuthal angle $\Phi$ between the
transverse pair momentum $\bbox{K}_\perp$ and the impact parameter 
$\bbox{b}$. Only these 6 combinations of the 10 independent space-time
variances $\langle \tilde{x}_{\mu} \tilde{x}_{\nu}\rangle$ can be
measured. 

For {\em azimuthally symmetric} collision ensembles the
emission function has a reflection symmetry $x_s\to-x_s$, eliminating 
3 of the 10 space-time variances, and the correlator is symmetric
under $q_s\to-q_s$ \cite{CNH95}. Then $R_{os}^2 = R_{sl}^2 =0$, and
the correlator is fully characterized by {\em 4 functions of only two
variables} $K_\perp$ and $Y$:
 \begin{equation}
 \label{PBC}
   C(\bbox{q},\bbox{K}) = 1 \pm \exp\left[-R_s^2 q_s^2 
   -R_o^2 q_o^2-R_l^2 q_l^2
   -2R_{ol}^2 q_o q_l\right],
 \end{equation}
with
 \begin{eqnarray}
   R_s^2(K_\perp,Y) &=& \langle \tilde x_s^2\rangle\, ,
 \qquad\qquad
   R_o^2(K_\perp,Y) = \langle (\tilde x_o - \beta_\perp\tilde t)^2 \rangle\, , 
 \nonumber\\
   R_l^2(K_\perp,Y) &=& \langle (\tilde x_l - \beta_l\tilde t)^2 \rangle\, , 
 \quad
   R_{ol}^2(K_\perp,Y) = \langle (\tilde x_o - \beta_\perp \tilde t)
          (\tilde x_l - \beta_l \tilde t)\rangle\, . 
 \label{PBCradii}
 \end{eqnarray}
These ``HBT radii'' mix spatial and temporal information on the source
in a non-trivial way, and their interpretation depends on the frame in
which the relative momenta $q$ are specified. Extensive discussions of
these parameters (including the cross-term $R_{ol}^2$ which originally
appeared in the important, but widely neglected paper \cite{P83}, was
recently rediscovered \cite{CSH95b} and then experimentally
confirmed \cite{AlberQM95,M96}) can be found in Refs. 
\cite{H97,CSH95a,CSH95b,CNH95,AS95,CL95,WSH96,HTWW96,T99,CSH95c,H96,TH98,SSX98}.
The cross-term vanishes in any longitudinal reference frame in which
the source is symmetric under $x_l\to-x_l$ \cite{P83} (e.g. for pion 
pairs with vanishing rapidity in the center-of-momentum system (CMS) of 
a symmetric collision); in general it does {\it not} vanish
for pion pairs with non-zero CMS rapidity, not even in the
longitudinally co-moving system (LCMS \cite{CP91}) \cite{CSH95c}.  

For azimuthally symmetric collisions no direction is distinguished
for pairs with $K_\perp=0$. As long as for $K_\perp\to 0$ the emission
function reduces to an azimuthally symmetric expression (an exception
is a certain class of opaque source models discussed in section
\ref{sec2.3.4}), one has at $K_\perp=0$ the identities \cite{CNH95}
$\langle \tilde x_o^2 - \tilde x_s^2 \rangle = \langle \tilde x_o \tilde
x_l\rangle = \langle \tilde t \tilde x_o\rangle =0$; these imply that
$R_o^2-R_s^2$ and the cross-term $R_{ol}^2$ vanish at $K_\perp=0$. At
non-zero $K_\perp$ these identities for the space-time variances may
be broken by transverse position-momentum correlations in the source,
as e.g. generated by transverse collective flow. If the latter are
sufficiently weak the leading $K_\perp$-dependence of the difference 
 \begin{equation}
 \label{Rdiff}
   R_{\rm diff}^2 \equiv R_o^2-R_s^2 = 
   \beta_\perp^2 \langle \tilde t^2\rangle 
   - 2 \beta_\perp \langle \tilde x_o\tilde t\rangle + 
   \langle \tilde x_o^2 - \tilde x_s^2\rangle
 \end{equation}
is given by the explicit $\beta_\perp$-dependence of the first term on 
the rhs. This yields the duration of the particle emission process 
$\Delta t = \sqrt{\langle t^2\rangle -\langle t\rangle^2}$
for particles with small $K_\perp$ \cite{B89,PCZ90,CP91}. (This is 
sometimes loosely called the ``lifetime'' of the effective source, but 
should not to be confused with the total time duration between nuclear 
impact and freeze-out which is not directly measurable.) 

The possibility to extract the emission duration from correlation 
measurements, pointed out by Bertsch and Pratt \cite{B89,PCZ90,CP91},
provided the main motivation for the construction of second generation 
experiments to measure high quality, high statistics correlation 
functions. Subsequent model studies for relativistic heavy-ion 
collisions \cite{fields,T99,ykp,TH98b,TH98c} where the emission 
duration is expected to be relatively short (of the order of the 
transverse source extension) showed, however, that the extraction of 
$\Delta t$ is somewhat model dependent; the relative smallness of the 
last two terms in (\ref{Rdiff}) cannot always be guaranteed, and their 
implicit $K_\perp$-dependence can mix with the explicit one of the 
interesting first term. 
 
The {\bf Yano-Koonin-Podgoretski\u{\i}\ (YKP) parametrization} is an
alternative Gaussian parametrization of the correlator for 
{\it azimuthally symmetric} collisions. It uses the mass-shell
constraint (\ref{ortho}) to express (\ref{3.4}) in terms of $q_\perp =
\sqrt{q_o^2+q_s^2},\, q_l$ and $q^0$ \cite{YK78,P83,CNH95,HTWW96}:
 \begin{equation}
 \label{YKP}
   C(\bbox{q},\bbox{K}) = 1 \pm \exp\Bigl[ - R_\perp^2 q_\perp^2
   - R_\parallel^2\Bigl(q_l^2-(q^0)^2\Bigr) 
   - \Bigl(R_0^2+R_\parallel^2\Bigr) (q\cdot U)^2\Bigr]\, .
 \end{equation}
Like (\ref{PBC}) it has 4 $(K_\perp,Y)$-dependent fit parameters: the
three radius parameters $R_\perp(\bbox{K})$, $R_\parallel(\bbox{K})$,
$R_0(\bbox{K})$, and a velocity parameter $U(\bbox{K})$ with only a
longitudinal spatial component:
  \begin{equation}
  \label{U}
    U(\bbox{K}) = \gamma(\bbox{K})\bigl(1,0,0,v(\bbox{K})\bigr),
    \qquad
    \gamma = (1-v^2)^{-1/2}.
  \end{equation}
The advantage of fitting the form (\ref{YKP}) to data is that the
extracted YKP radii $R_\perp, R_\parallel, R_0$ do not depend on the
longitudinal velocity of the measurement frame, while the fourth fit
parameter $v(\bbox{K})$ is simply boosted by that velocity. The frame
in which $v(\bbox{K})=0$ is called the Yano-Koonin (YK) frame; the YKP 
radii are most easily interpreted in terms of coordinates measured in 
this frame \cite{CNH95}:
 \begin{eqnarray}
   R_\perp^2(\bbox{K}) &=& R_s^2(\bbox{K}) 
                         = \langle \tilde x_s^2\rangle\, ,
 \label{YKPradii1}\\
   R_\parallel^2(\bbox{K}) &=& 
   \langle (\tilde x_l-(\beta_l/\beta_\perp)\tilde x_o)^2\rangle
   -(\beta_l/\beta_\perp)^2 \langle\tilde x_s^2\rangle 
   \approx \langle \tilde x_l^2 \rangle\, ,
 \label{YKPradii2}\\
   R_0^2(\bbox{K}) &=& 
   \langle (\tilde t-\tilde x_o/\beta_\perp)^2\rangle
   -\langle\tilde x_s^2\rangle/\beta_\perp^2 
   \approx \langle \tilde t^2 \rangle\, ,
 \label{YKPradii3}
 \end{eqnarray}
where the approximations in the last two lines are equivalent to
dropping the last two terms in (\ref{Rdiff}) (see discussion above).
To the extent that these hold, the three YKP radii thus have a
straightforward interpretation as the transverse, longitudinal and
temporal homogeneity lengths in the YK frame. In particular the time
structure of the source only enters in $R_0$. For sources with strong
longitudinal expansion, like those created in relativistic heavy-ion
collisions, it was shown in extensive model studies
\cite{HTWW96,T99,ykp,TH98b,TH98c} that the YK velocity $v(\bbox{K})$  
very accurately reflects the longitudinal velocity at the center
$\bar x(\bbox{K})$ of the ``homogeneity region'' of particles of
momentum $\bbox{K}$. The YK frame can thus be interpreted as the rest
frame of the effective source of particles with momentum $\bbox{K}$,
and the YKP radii measure the transverse, longitudinal and temporal
size of this effective source in its own rest frame.

The parametrizations (\ref{PBC}) and (\ref{YKP}) use different
independent components of $q$ but are mathematically equivalent. The
YKP parameters can thus be calculated from the Cartesian ones and vice
versa \cite{HTWW96}. The corresponding relations are  
 \begin{eqnarray}
 \label{24}
   R_s^2 &=& R_\perp^2\, ,
 \\
 \label{24a}
   R_{\rm diff}^2 &=& R_o^2 - R_s^2 = \beta_\perp^2 \gamma^2 
             \left( R_0^2 + v^2 R_\parallel^2 \right) \, ,
 \\
 \label{24b}
   R_l^2 &=& \left( 1 - \beta_l^2 \right) R_\parallel^2 
             + \gamma^2 \left( \beta_l-v \right)^2
             \left( R_0^2 + R_\parallel^2 \right) \, ,
 \\
 \label{24c}
   R_{ol}^2 &=& \beta_\perp \left( -\beta_l R_\parallel^2 
             + \gamma^2 \left( \beta_l-v \right)
             \left( R_0^2 + R_\parallel^2 \right) \right) \, ,
 \end{eqnarray}
whose inversion reads
 \begin{eqnarray}
 \label{22a}   
   R_\parallel^2 = B - v\, C ,&&\!\!
   R_0^2 = A - v\, C ,\quad
   v = {A+B\over 2C} \left( 1 - \sqrt{1 - \left({2C\over A+B}\right)^2}
                       \right) ,
 \\
 \label{21a}
   A = {1\over \beta_\perp^2}\, R_{\rm diff}^2,&&\!\!\!\!\!\!
   B = R_l^2{-}2 {\beta_l\over\beta_\perp} \, 
       R_{ol}^2{+}{\beta_l^2\over\beta_\perp^2} \, R_{\rm diff}^2 ,\ \ 
   C = -{1\over\beta_\perp} \, 
       R_{ol}^2{+}{\beta_l\over\beta_\perp^2} R_{\rm diff}^2 . 
 \end{eqnarray}
These last definitions hold in an arbitrary longitudinal reference
frame. According to (\ref{22a}) $v$ is zero in the frame where $C$
vanishes. However, (\ref{22a}) also shows that the YKP
parametrization becomes ill-defined if the argument of the square root
turns negative. This can indeed happen, in particular for opaque
sources \cite{HV98,TH98b}; this has motivated the introduction of a {\em
  modified YKP parametrization} in \cite{T99,TH98b,TH98c} which avoids
this problem at the expense of a less intuitive interpretation of the
modified YKP radii. These remarks show that these relations provide an 
essential check for the internal consistency of the Gaussian fit to
the measured correlation function and for the physical interpretation
of the resulting HBT parameters.


\subsection{Collective Expansion and $\protect\bbox{K}$-Dependence of
  the Correlator} 
\label{sec2.3}

If the particle momenta are correlated with their emission points
(``$x$-$p$-corre\-la\-tions''), the space-time variances in (\ref{3.4})
depend on the pair momentum $\bbox{K}$. Various mechanisms can lead to
such correlations; the most important one for heavy-ion collisions is
collective expansion of the source. Recently a major effort has been
launched to extract the collective flow pattern at freeze-out from the
$\bbox{K}$-dependence of the HBT parameters. However, thermalized
sources may exhibit temperature gradients along the freeze-out
surface which cause additional $x$-$p$-correlations. Moreover, 
pion spectra receive sizeable contributions from the decay of unstable
resonances some time after freeze-out. These decay pions tend to come
from a somewhat larger space-time region than the directly emitted
ones and, due to the decay kinematics, they preferentially populate
the low-momentum region. Together these two effects also generate
$x$-$p$-correlations for the emitted pions even if the original source
did not have them \cite{GP88}.

The separation of these different effects requires extensive model
studies some of which will be reviewed below. For didactical reasons 
we will discuss them in the context of the YKP parametrization where 
certain mechanisms can be demonstrated most transparently. A 
translation for the Cartesian fit parameters via the cross-check 
relations (\ref{24})-(\ref{24c}) is straightforward. Furthermore we 
will show that for sources with strong longitudinal 
expansion the YK frame (effective source rest frame) is usually rather 
close to the LCMS (in which the pairs have vanishing longitudinal 
momentum), {\it i.e.} $v(\bbox{K})\approx 0$ in the LCMS. This allows, 
at least qualitatively, for a rather direct extraction of source 
properties in its own rest frame from the Cartesian HBT radius 
parameters in the LCMS. This is important since initially most 
multi-dimensional HBT analyses were done with the Cartesian 
parametrization in the LCMS, before the YKP parametrization became 
popular.

We will concentrate on the discussion of azimuthally symmetric sources 
(central collisions) for which extensive knowledge, both theoretical 
and experimental, has been accumulated in the last few years. Many 
analytical model studies
\cite{CSH95a,CSH95b,CNH95,CL95,AS95,S95,WSH96,HTWW96,T99,CSH95c,TH98,ykp,TH98b,TH98c,CN96,D99}
are based on the following parametrization of the emission function
(or slight variations thereof):  
 \begin{equation}
 \label{3.15}
    \!\!\!S(x,K)\! =\! {M_\perp \cosh(\eta{-}Y) \over 8 \pi^4 \Delta \tau}
    \exp\!\left[- {K{\cdot}u(x) \over T(x)}
                       - {(\tau{-}\tau_0)^2 \over 2(\Delta \tau)^2}
                       - {r^2 \over 2 R^2(\eta)} 
                       - {(\eta{-}\eta_0)^2 \over 2 (\Delta \eta)^2}
         \right].
 \end{equation}
Here $r^2{=}x^2{+}y^2$, $\eta{=}{1 \over 2} \ln[(t{+}z)/(t{-}z)]$, and 
$\tau{=}(t^2{-}z^2)^{1/2}$ parametrize the space-time coordinates
$x^\mu$, with $d^4x = \tau\, d\tau\, d\eta\, r\, dr\, d\phi$.
$Y{=}{1\over 2} \ln[(E_K{+}K_L)/(E_K{-}K_L)]$ and
$M_\perp{=}(m^2{+}K_\perp^2)^{1/2}$ parametrize the pair momentum
$\bbox{K}$. $\sqrt{2} R$ is the 2-di\-men\-sional transverse rms radius 
of the source ($\langle r^2\rangle = 2R^2$); $R$ is usually taken as a 
constant. $\tau_0$ is the average freeze-out proper time, $\Delta \tau$ 
the mean duration of particle emission, and $\Delta \eta$ controls 
the longitudinal size of the source, $L\sim\tau_0\sinh\eta$. 
Note that $S(x,K)$ describes the phase-space distribution at 
freeze-out and not the dynamical {\em evolution} of 
the source from initial conditions; the latter is described by 
hydrodynamical or microscopic kinetic models, discussed below. 

The Boltzmann factor $\exp[-K{\cdot}u(x)/T(x)]$ parametrizes the 
momen\-tum-space structure of the source in terms of a collective, 
directed component, given by a flow velocity field $u^\mu(x)$, and a 
randomly fluctuating component, characterized by an exponential spectrum 
with local slope $T(x)$, as suggested by the shape of the measured 
single-particle spectra. Although this parametrization is
somewhat restrictive because it implies that the random component is
locally isotropic, it does not {\em require} thermalization of the
source at freeze-out. But if it turns out that all particle species
can be described {\em simultaneously} by the emission function
(\ref{3.15}), with the {\em same} temperature and velocity fields
$T(x)$, $u(x)$, this would indeed suggest thermalization as the most
natural explanation.  

It is convenient to decompose $u^\mu(x)$ in the form 
 \begin{equation}
 \label{26}
   u^\mu(x) = \left( \cosh \eta_l \cosh \eta_t, \,
                     \sinh \eta_t\, \bbox{e}_r,  \,
                     \sinh \eta_l \cosh \eta_t \right) ,
 \end{equation}
with longitudinal and transverse flow rapidities $\eta_l(x)$ and
$\eta_t(x)$. A simple boost-invariant longitudinal flow
$\eta_l(\tau,\eta,r)=\eta$ ($v_l = z/t$) is commonly assumed. For
the transverse flow rapidity profile the simplest choice is  
 \begin{equation}
 \label{27}
  \eta_t(\tau,\eta,r) = \eta_f(\tau,\eta) 
  \left( {r \over R(\eta)} \right)
 \end{equation} 
with a scale parameter $\eta_f$. (In our notation, $\eta_f$ denotes 
the transverse collective flow rapidity, whereas $\beta_\perp$ is the 
transverse velocity of the particle pair.) In most studies $\eta_f$ was 
(like $R$) set constant such that $\eta_t(r)$ was a function of $r$ only. 
This cannot reproduce the observed rapidity dependence of $\langle
p_\perp\rangle$ and of the inverse slopes of the $m_\perp$-spectra 
\cite{AGSspectra,NA49spectra}. In \cite{CN96,D99} it was shown that
an $\eta$-dependence of $R$, with $R(\eta)$ shrinking in the backward
and forward rapidity regions keeping the slope $\eta_f/R$ in 
(\ref{27}) fixed, is sufficient to fully repair this deficiency. Here 
we discuss only the simpler case of constant $R,\, \eta_f$. 

With these assumptions the exponent of the Boltzmann factor in
(\ref{3.15}) becomes  
 \begin{equation}
 \label{26a}
  K\cdot u(x) = M_\perp \cosh(Y-\eta) \cosh\eta_t(r) - 
                \bbox{K}_\perp{\cdot}\bbox{e}_r \sinh\eta_t(r)\, .
 \end{equation}
For vanishing transverse flow ($\eta_f=0$) the source depends only 
on $M_\perp$ and remains azimuthally symmetric for all $K_\perp$.
%
\begin{figure}[ht]
\vspace*{11cm}
\includegraphics{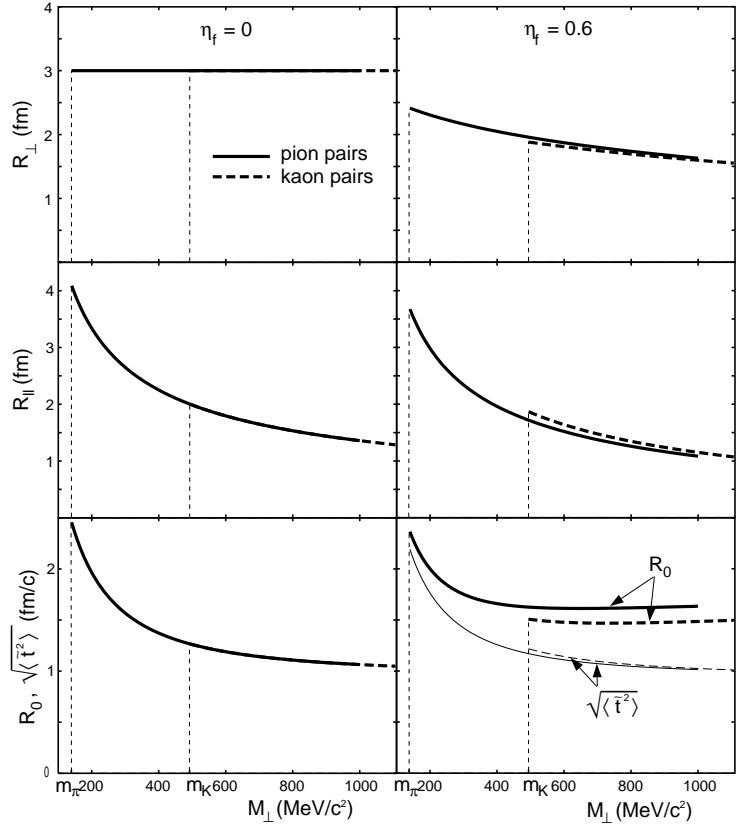}
\caption{The YKP radii $R_\perp$, $R_\parallel$, and $R_0$ (top to
  bottom) as functions of $M_\perp$ for pairs at $Y_{\rm cm}=0$. Left:
  no transverse flow. Right: $\eta_f=0.6$. Solid (dashed) lines are
  for pions (kaons). Note the breaking of $M_\perp$-scaling by
  transverse flow. Source parameters: $T=140$ MeV, $\Delta\eta=1.2$,
  $R=3$ fm, $\tau_0=3$ fm/$c$, $\Delta\tau=1$ fm/$c$. (Figure taken 
  from \protect\cite{H96}.)
  \label{F1}}
\end{figure} 
%
Since in the absence of transverse flow the $\beta$-dependent terms in 
(\ref{YKPradii2}) and (\ref{YKPradii3}) vanish and the source itself
depends only on $M_\perp$, all three YKP radius parameters then show
{\em perfect $M_\perp$-scaling}: plotted as functions of $M_\perp$,
they coincide for pion and kaon pairs (see Figure~\ref{F1}, left
column). This remains true if $T(x)$ varies with $x$; temperature
gradients in the source do not destroy the $M_\perp$-scaling.

For $\eta_f\ne 0$ (right column) this $M_\perp$-scaling is broken by
two effects: (1) The thermal exponent (\ref{26a}) receives an
additional contribution proportional to $K_\perp$. (2) The terms which
were neglected in the approximations (\ref{YKPradii2},\ref{YKPradii3})
are non-zero, and they also depend on $\beta_\perp=K_\perp/E_K$. Both 
induce an explicit rest mass dependence and destroy the
$M_\perp$-scaling of the YKP size parameters. 


\subsubsection{Longitudinal flow: Yano-Koonin rapidity and
  $M_\perp$-dependence of $R_\parallel$}
\label{sec2.3.1}

At each point in an expanding source the local velocity distribution
is centered around the average fluid velocity $u(x)$. Thus two fluid
elements moving rapidly relative to each other are unlikely to 
contribute particles with small relative momenta. Only source regions 
moving with velocities close to that of the observed particle pair 
contribute to the correlation function. How close is controlled by 
the width of the random component in the momentum distribution: the 
larger the local ``thermal smearing'', the more the differences in the 
fluid velocities can be balanced out, and the larger the 
``regions of homogeneity'' in the source become.

\begin{figure}[ht]
\vspace*{5cm}
\includegraphics{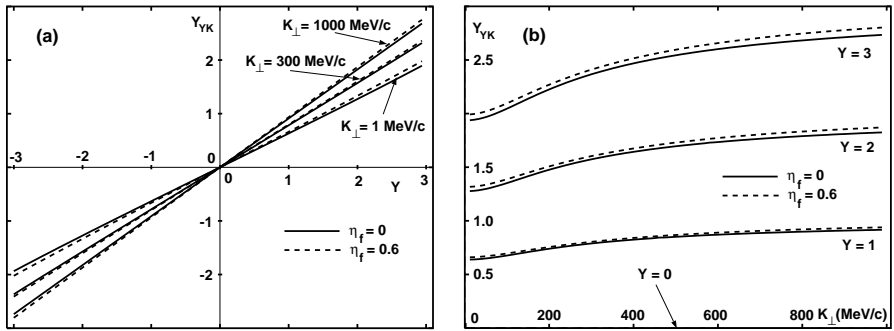}
\caption{(a) The Yano-Koonin rapidity for pion pairs, as a function of
  the pair c.m. rapidity $Y$, for various values of $K_\perp$ and two
  values for the transverse flow $\eta_f$. (b) The same, but plotted
  against $K_\perp$ for various values of $Y$ and $\eta_f$. Source
  parameters as in Fig.~\protect\ref{F1}. 
  (Figure taken from \protect\cite{HTWW96}.)
\label{F2}} 
\end{figure}

Longitudinal expansion is most clearly reflected in the behaviour of the 
Yano-Koonin (YK) rapidity $Y_{_{\rm YK}}{=}\frac 12\ln[(1{+}v)/(1{-}v)]$.
Figure~\ref{F2} shows (for pion pairs) its dependence on the pair
momentum $\bbox{K}$. Transverse flow is seen to have a negligible
influence on the YK rapidity. On the other hand, the linear dependence 
of $Y_{_{\rm YK}}$ on the pair rapidity $Y$ (Figure~\ref{F2}a) is a 
direct reflection of the longi\-tu\-dinal expansion flow \cite{HTWW96}; 
for a non-expanding source $Y_{_{\rm YK}}$ would be independent of $Y$. 
The correlation between the velocities of the pair ($Y$) and of the 
emission region ($Y_{_{\rm YK}}$) strengthens as the thermal smearing 
decreases. For the Boltzmann form (\ref{3.15}) the latter is controlled 
by $T/M_\perp$, and correspondingly $Y_{_{\rm YK}}(K_\perp,Y) \to Y$ 
as $K_\perp\to\infty$. For small $K_\perp$, thermal smearing weakens 
the correlation, and the effective source moves somewhat more slowly 
than the observed pairs, whose longitudinal velocities have an 
additional thermal component. 

If the ratio of the longitudinal source velocity gradient to the 
thermal smearing factor, defined in (\ref{LH}) below, is large, 
$R_\parallel$ becomes small and the longitudinal rapidity $Y_{_{\rm YK}}$
of the effective source becomes equal to that of the emitted pairs, $Y$.
This can be true even for slow longitudinal expansion as long as it is 
strong enough compared to the thermal smearing. Consequently, observation 
of a behaviour like the one shown in Figure~\ref{F2}a demonstrates 
{\em strong}, but not necessarily {\em boost-invariant} longitudinal 
flow.  

Longitudinal expansion is also reflected in the $M_\perp$-dependence
of $R_\parallel$ (second row of Figure~\ref{F1}). Comparison of the left
with the right diagram shows only minor effects from transverse
expansion; longitudinal and transverse dynamics are thus cleanly
separated. A qualitative understanding of the $M_\perp$-dependence
is provided by the following expression, valid for pairs with
$Y=0$, which can be derived by evaluating (\ref{YKPradii2}) via
saddle-point integration \cite{MS88,CSH95a,AS95,CL95}: 
 \begin{eqnarray}
 \label{Lstar}
   &&R_\parallel^2 \approx L_*^2 \equiv {L_{\rm flow}^2 \over 
   1 + \left(L_{\rm flow}/\tau_0\Delta\eta\right)^2}\, , 
 \\
 \label{LH}
   &&L_{\rm flow}(M_\perp) 
   = {1\over \partial{\cdot}u_l} \, \sqrt{{T\over M_\perp}}
   = \tau_0\, \sqrt{{T\over M_\perp}}\, ,
 \end{eqnarray}
Eq.~(\ref{LH}) shows explicitly the competition between the
longitudinal velocity gradient $\partial{\cdot}u_l$ and the thermal
smearing factor $T/M_\perp$. For {\em strong} longitudinal expansion
(large velocity gradient and/or weak thermal smearing) and/or large
geometric longitudinal extension $\tau_0\Delta\eta$ of the source the
second term in the denominator can be neglected, and $R_\parallel$
drops steeply as $1/\sqrt{M_\perp}$ \cite{MS88}. Note that
quantitative corrections to (\ref{Lstar}) are not always small 
\cite{WSH96}. 

We emphasize that only the first equation in (\ref{LH}) is general. 
The appearance of the parameter $\tau_0$ in the second equation is
due to the choice of a Bjorken profile for the longitudinal flow for
which the longitudinal velocity gradient is given by the total proper
time $\tau_0$ between impact and freeze-out. This is not true in
general; the interpretation of the {\rm length} $R_\parallel$ in 
terms of the total expansion {\rm time} is therefore a highly 
model-dependent procedure which should be avoided. As a matter of 
principle, the absolute temporal position of the freeze-out point 
is not measurable, see section \ref{sec2.2.1}.    
 
Strong longitudinal $x$-$p$-correlations also occur in string
fragmentation. In fact, in the Schwinger model of string breaking 
\cite{string} the quark pairs created from the chromoelectric field
are assumed to have longitudinal momentum $Y=\eta$, without thermal
fluctuations. Thus a similar linear rise of the YK-rapidity with the 
pair rapidity and a strong decrease of $R_\parallel(M_\perp)$ would
also be expected in jet fragmentation (with the $x_l$-axis oriented
along the jet axis). It would be interesting to confirm this
prediction \cite{ykp} in $e^+e^-$ or $pp$ collisions.


\subsubsection{Transverse flow: $M_\perp$-dependence of $R_\perp$}
\label{sec2.3.2}

Just as longitudinal expansion affects $R_\parallel$, transverse flow
causes an $M_\perp$-depen\-dence of $R_\perp$. This is seen in the first
row of diagrams in Figure~\ref{F1}, which also shows that longitudinal
flow does not contribute to this feature: for $\eta_f=0$ the
transverse radius does not depend on $M_\perp$, in spite of strong
longitudinal expansion of the source. A qualitative understanding of
this behaviour is given by the analogue of (\ref{Lstar}), again
obtained by evaluating (\ref{YKPradii1}) via saddle point integration
\cite{CSH95a,CL95,CNH95}: 
 \begin{eqnarray}
 \label{Rstar}
   &&R_\perp^2 \approx R_*^2 \equiv {R^2 \over 
   1 + \left(R/R_{\rm flow}\right)^2} = 
   {R^2 \over 1 + \eta_f^2 (M_\perp/T)}\, , 
 \\
 \label{RH}
   &&R_{\rm flow}(M_\perp) 
   = {1\over \partial \eta_t(r)/\partial r} \, \sqrt{{T\over M_\perp}}
   = {R\over \eta_f}\, \sqrt{{T\over M_\perp}}\, .
 \end{eqnarray}
Once again there is competition between flow velocity gradients in
the source, this time in the transverse direction, which tend to
reduce the homogeneity regions, and thermal smearing by the factor
$T/M_\perp$ resulting from the random component in the momentum
distribution, enlarging the regions of homogeneity. The
left equations in (\ref{Rstar},\ref{RH}) are generic while the right
ones apply to the specific transverse flow profile (\ref{27}). 

Transverse flow must be built up from zero during the collision
while longitudinal $x$-$p$-correlations may contain a sizeable
primordial component from incomplete stopping of the two nuclei and/or 
the particle production process (e.g. string fragmentation, see
above). One thus expects generically weaker transverse than
longitudinal flow effects at freeze-out. Correspondingly, in 
realistic simulations (e.g. \cite{CL95,ykp,fields}) the longitudinal 
homogeneity length $R_\parallel$ turns out to be dominated by the 
expansion ({\it i.e.} by $L_{\rm flow}$) while in the transverse 
direction the geometric size $R$ dominates at small $M_\perp$, with 
flow effects taking over only at larger values of $M_\perp$. 
Correspondingly, $R_\perp(M_\perp)$ decreases more slowly at small 
$M_\perp$ than $R_\parallel(M_\perp)$ \cite{WSH96,ykp}. 


\subsubsection{The emission duration}
\label{sec2.3.3}

Saddle-point integration of (\ref{YKPradii3}) with the source
(\ref{3.15}) yields, with $L_*$ from (\ref{Lstar}),
 \begin{equation}
 \label{Tstar}
   R_0^2 \approx (\Delta t_*)^2 \equiv (\Delta\tau)^2 + 
   2 \left( \sqrt{\tau_0^2 + L_*^2} - \tau_0 \right)^2 \, .
 \end{equation}
The $M_\perp$-dependence of $L_*$ thus induces an $M_\perp$-dependence 
of the temporal YKP parameter. Eq.~(\ref{Tstar}) reflects the proper 
time freeze-out assumed in the model (\ref{3.15}): particles emitted 
at different points $z$ are also emitted at different global times $t$, 
and the total temporal width of the effective source is thus given by 
the Gaussian width $\Delta\tau$ plus the additional variation along the 
proper time hyperbola, integrated over the longitudinal homogeneity 
region $L_*$ \cite{WSH96}.

Although (\ref{Tstar}) suggests that the proper emission duration
$\Delta\tau$ can be measured via $R_0$ in the limit $M_\perp\to\infty$ 
(where $L_*\to 0$), this is has not been true in systems studied to date.
For transversely expanding sources $R_0$ receives additional contributions
from the $\beta_\perp$-dependent terms in (\ref{YKPradii3}), in
particular at large $M_\perp$ (see Figure~\ref{F1}). The most important 
correction is due to the term 
$\langle \tilde x_o^2-\tilde x_s^2\rangle/\beta_\perp^2$ which
can have either sign and usually grows with $M_\perp$
\cite{ykp,TH98c,HV98}. The extraction of the emission duration 
must thus be considered the most model-dependent aspect of the 
HBT analysis.   


\subsubsection{Temperature gradients and opacity effects}
\label{sec2.3.4}

A different source of transverse $x$-$p$-correlations which can
compete with transverse flow in generating an $M_\perp$-dependence 
of $R_\perp$ are transverse temperature gradients in the emission
function. Since particle densities and mean free paths (which control
the freeze-out process \cite{BGZ78,LH89}) depend very strongly on
temperature, one would {\em a priori} not expect strong temperature
variations across the freeze-out surface \cite{MH97}. While transverse 
temperature gradients and transverse flow affect $R_\perp$ similarly 
(except that the latter weakly breaks the $M_\perp$-scaling), they have 
quite different effects on the single particle spectra \cite{CL95,TH98}: 
transverse temperature gradients strongly reduce the flattening effect of
transverse flow on the $m_\perp$-spectra which is needed to reproduce 
the data \cite{LH89,NA44flow}. Thus constrained by single-particle data, 
their phenomenological usefulness is limited. Temporal temperature 
gradients only reduce the emission duration, but do not affect the 
transverse $R$-parameter \cite{CL95,TH98}.

One possible source feature that parametrization
(\ref{3.15}) cannot describe is ``opacity'', {\it i.e.} 
surface dominated emission. Heiselberg and Vischer generated 
opaque sources by multiplying (\ref{3.15}) (or a similar
source with bulk freeze-out) with an exponential absorption factor
\cite{HV98} (see also \cite{TH98b}) 
 \begin{eqnarray}
 \label{opaque}
      S_{\rm opaque}(x,p) &=& S(x,p)\, 
         \exp\left[-\sqrt{8/\pi} 
         \left( l_{\rm eff}/\lambda_{\rm mfp}\right) \right],
 \\
      l_{\rm eff} &=& l_{\rm eff}(r,\phi)
      = e^{-{x_s^2\over 2R^2}} \int_{x_o}^\infty
        e^{-{x'^2\over 2R^2}}\, dx'\, .
 \label{5.10}
 \end{eqnarray}
The ratio $\lambda_{\rm mfp}/R$ controls the degree of opacity of the
source; as $\lambda_{\rm mfp}/R\to 0$, the source becomes an
infinitely thin radiating shell. The parametrization (\ref{opaque})
together with (\ref{3.15}) leads to sources with negative 
$\langle \tilde x_o^2 -\tilde x_s^2 \rangle$ {\em for all values} of
$K_\perp$ (including the limit $K_\perp\to 0$). According to
(\ref{Rdiff}) and (\ref{YKPradii3}) this leads to negative values
of $R_{\rm diff}^2$ and $R_0^2$ ($R_0^2$ even diverges as $K_\perp\to
0$ \cite{TH98b}); the data (see below) are consistent with vanishing or
positive $R_{\rm diff}^2$ at small $K_\perp$.

A non-vanishing difference $\langle \tilde x_o^2 -\tilde x_s^2 \rangle$ in 
the limit $K_\perp\to 0$ violates the postulated azimuthal symmetry of 
the source (see discussion before Eq.~(\ref{Rdiff})). It is easy
to see that short-lived sources can never be opaque for particles with
$K_\perp\to 0$: the source shrinks to zero before such particles can
be reabsorbed. The particular behaviour excluded in \cite{TH98b} 
is thus anyhow rather unphysical. At larger $K_\perp$, on the other
hand, the ``opacity signal'' $\langle \tilde x_o^2{-}\tilde 
x_s^2 \rangle<0$ (leading, if strong enough, to $R_o^2 < R_s^2$ 
\cite{HV98}) can be ``faked'' by other mechanisms: Tom\'a\v{s}ik found 
\cite{T99,TH98c} that expanding sources with a box-like transverse 
density profile generate exactly such a signature. At the moment it is 
thus unclear how to uniquely distinguish ``opaque'' from ``transparent'' 
sources.


\subsection{Non-Gaussian Features of the Correlator and $q$-Moments}
\label{sec2.4}

The ($K$-dependent) HBT radii provide a full characterization of the
two-particle correlation function only if it is a Gaussian in $q$ or,
equivalently, if the effective source $S(x,K)$ is a Gaussian in
$x$. However, in many physical situations the source is not
characterized by just one, but by several distinct length scales. In
this case the Gaussian approximation of section \ref{sec2.2} breaks
down. 

\subsubsection{Resonance decays}
\label{sec2.4.1}

The most important physical processes leading to a
non-Gaussian shape of the correlator are resonance decays 
\cite{GP88,schlei,Heis96,SOPSW96,CLZ96,WH97,H97}. Especially longlived
resonances which decay into pions cause a long-range exponential tail
in the pion emission function which distorts the two-particle
correlation function at small relative momentum $q$ (see example
in Figure 4 below). According to
\cite{WH97,sulli} the resonances can be classified into three classes: 

\begin{enumerate}
 
\item 
{\em Short-lived resonances} ($\Gamma>30$ MeV) which (especially if
heavy) decay very close to their production point. Their most
important effect is to add a contribution proportional to their 
lifetime to the emission duration \cite{Heis96,WH97}, thereby
affecting $R_l^2$ and $R_o^2$ in the Cartesian and $R_0^2$ in the YKP 
parametrization, but not the transverse radius $R_\perp$. They do not
spoil the Gaussian parametrization. 
 
\item
{\em Long-lived resonances} ($\Gamma\ll 1$ MeV), mostly the
$\eta,\eta',K_S^0$ and hyperons. These resonances travel far outside
the original source before decaying. The resulting wide tail in the
emission function contributes to the correlator only at very small
relative momenta. This region is experimentally inaccessible due to
finite two-track and momentum resolution, and the contribution from
this ``halo'' \cite{CLZ96} to the correlator is thus missed in the
experiment. The result is an apparent decrease of the correlation
strength ({\it i.e.} the intercept $\lambda$). In the measurable
$q$-range the {\em shape} of the correlator is not affected.  

\item
{\em The $\omega$ meson} ($\Gamma=8.4$ MeV) is not sufficiently
longlived to escape detection in the correlator. Its lifetime is, 
however, long enough to create a measurable exponential tail in the 
pion emission function which distorts the shape of the correlator, 
giving it extra weight at small $q$ and destroying its Gaussian form.

\end{enumerate}

In practice the pions from short-lived resonances can thus be simply
added to the directly emitted ones into an emission function for the 
``core'' \cite{schlei,CLZ96}. The ``halo'' from long-lived resonances 
is accounted for by a reduced intercept parameter  
 \begin{equation}
 \label{lambda}
   \lambda(\bbox{K}) = \left( 1 - \sum_r f_r(\bbox{K})\right)^2,
 \end{equation}
where the sum goes over all longlived resonances and $f_r(\bbox{K})$
is the fraction of pions with momentum $\bbox{K}$ stemming from
resonance $r$. A correspondingly modified Cartesian parametrization 
for the correlator reads 
 \begin{equation}
 \label{modCart}
   C(\bbox{q},\bbox{K}) = 1 \pm \lambda(\bbox{K})
   \exp\left( - \sum_{ij} R_{ij}^2(\bbox{K}) q_i q_j\right).
 \end{equation}
Pions from $\omega$ decays must, however, be considered explicitly
and, if sufficiently abundant, the resulting correlator is no longer
well described by the ansatz (\ref{modCart}).   

In heavy-ion collisions the resonance fractions $f_r$ are unknown
since most resonances cannot be reconstructed in the
high-multiplicity environment. Thus $\lambda(\bbox{K})$ in
(\ref{modCart}) is an additional fit parameter. Its value is very
sensitive to non-Gaussian distortions in the correlator, and so are
the HBT radii extracted from a fit to the function (\ref{modCart}). In
theoretical studies \cite{WH97} it was found that differences of more
than 1 fm in the fitted HBT radii can occur if the fit is performed
with $\lambda$ fixed to its theoretical value (\ref{lambda}) or if, as 
done in experiment, $\lambda$ is fitted together with the radii. In
the latter case resonance contributions (including the $\omega$)
affect the fitted radii much less than in the former. This difference
in procedure may largely explain the consistently larger resonance
contributions to the HBT radii found by Schlei et al. 
\cite{schlei,SOPSW96}, compared to the much weaker effects reported in
\cite{WH97}. Whereas Schlei et al. \cite{schlei,SOPSW96} find that
resonances, whose decay pions contribute only to the region of small
$K_\perp$, add considerably to the $M_\perp$-dependence of
$R_s=R_\perp$ and thus contaminate the transverse flow signature,
practically no such effect was found in \cite{WH97}. 


\subsubsection{$q$-moments}
\label{sec2.4.2}

In view of these systematic uncertainties one may ask for a more
quantitative characterization of correlation functions whose shape
deviates from a Gaussian. This can be achieved via the so-called
$q$-moments of the correlator \cite{WH97}. In this approach the matrix 
of Cartesian HBT parameters ${\cal R} \equiv (R^2_{ij}(\bbox{K}))\
(i,j=o,s,l)$ and the correlation strength $\lambda(\bbox{K})$ are
calculated from the following integrals:
 \begin{eqnarray}
 \label{4.7} 
   &&{\textstyle{1\over 2}} \left( {\cal R}^{-1}\right)_{ij} =
     \langle\!\langle q_i\, q_j \rangle\!\rangle(\bbox{K}) =
     {\int d^3q\,\, q_i\, q_j\, [C(\bbox{q},\bbox{K}) - 1]
      \over \int d^3q\, [C(\bbox{q},\bbox{K}) - 1]},
 \\
 \label{4.8} 
   &&\lambda(\bbox{K}) = \sqrt{\det {\cal R}(\bbox{K})/\pi^3}
        \int d^3q\,[C(\bbox{q},\bbox{K}) - 1].
 \end{eqnarray}
Similar expressions exist for the YKP parameters~\cite{WH97}. For a 
Gaussian correlator this gives the same HBT parameters as a Gaussian
fit; for non-Gaussian correlators the HBT radius parameters and
intercept are {\it defined} by (\ref{4.7},\ref{4.8}). 

Deviations of the correlator from a Gaussian shape are then quantified
by higher order $q$-moments. Formally they can be obtained as
derivatives at the origin of the relative source function
$S_{\bbox{K}}(\bbox{r})$ which acts as generating function \cite{WH97}:   
 \begin{equation}
 \label{4.10} 
    \langle\!\langle q_{i_1}\cdots q_{i_n}\rangle\!\rangle(\bbox{K})
    = {(-i\partial)^n\over \partial r_{i_1}\cdots \partial r_{i_n}} 
      \ln S_{\bbox{K}}(\bbox{r})\Big\vert_{\bbox{r}=0} \, .
 \end{equation}
Practical applications of this method are limited by severe statistical 
requirements for the measured correlator. So far they have been
restricted to uni-directional moments along one of the three $q$-axes. 
The leading deviation of $C(\bbox{q},\bbox{K})$ from a Gaussian shape
is then given by the {\em kurtosis} \cite{WH97} 
  \begin{equation}
  \label{4.16}
    \Delta_i(\bbox{K}) = 
    {\langle\!\langle q_i^4 \rangle\!\rangle
     \over 3 \langle\!\langle q_i^2 \rangle\!\rangle^2}(\bbox{K}) -
   1\, ,\quad i=o,s,l.
  \end{equation}
In \cite{WH97} the influence of transverse flow and resonance decays
on the transverse HBT radius $R_s=R_\perp$ on the kurtosis
$\Delta_s(K_\perp)$ was studied for sources of the type
(\ref{3.15}). It was found that decay pions give a positive
contribution to the kurtosis which disappears at large $K_\perp$
together with the resonance fractions $f_r(\bbox{K})$. Transverse
flow, on the other hand, leads to a vanishing or very small negative
kurtosis which tends to become larger with $K_\perp$. The sign and
$K_\perp$-dependence of the kurtosis thus provide a possibility to
check whether a measured $K_\perp$-dependence of $R_s$ is really due
to transverse flow or ``faked'' by resonance decays \cite{WH97}.  


\subsection{The Average Freeze-out Phase-Space Density}
\label{sec2.5}

Bertsch \cite{B94} pointed out that by combining measurements of
single-particle momentum spectra and two-particle correlations 
one can determine the spatially averaged phase-space density at
freeze-out and thereby test local thermal equilibrium in the pion
source created in high energy nuclear collisions:
 \begin{equation}
 \label{12}
   \langle f \rangle(p) = 
   \frac{\int_\Sigma f^2(x,p)\, p^\mu d^3\sigma_\mu(x)} 
        {\int_\Sigma f(x,p)\, p^\mu d^3\sigma_\mu(x)} .
 \end{equation}
Here $d^3\sigma(x)$ is the normal vector on a space-like space-time
hypersurface $\Sigma(x)$. According to Liouville's theorem, $\Sigma$
is arbitrary as long as its time arguments are later than the time
$t_{\rm f}(\bbox{x})$ at which the last pion passing the surface at
point $\bbox{x}$ was produced. If the measured single-particle
$p_\perp$-spectrum is parametrized by an exponential with inverse 
slope parameter $T_{\rm eff}(y)$ and the two-particle correlation
function by the Gaussian (\ref{PBC}), one finds \cite{M96,B94,WH99}
 \begin{equation}
 \label{psdens}
   \langle f \rangle (K_\perp,Y) = {\sqrt{\lambda(K_\perp,Y)}\, 
    (dn/dY)\, (2\pi T^2_{\rm eff}(Y))^{-1} \, 
    e^{-K_\perp/T_{\rm eff}(Y)} 
    \over 
    \pi^{-3/2}\, E_p\, R_s(K_\perp,Y) 
    \sqrt{R^2_o(K_\perp,Y)R^2_l(K_\perp,Y)-R^4_{ol}(K_\perp,Y)}}\, .
 \end{equation}
The numerator (where $dn/dY$ denotes the multiplicity density of a 
{\em single} charge state) gives the momentum-space density at 
freeze-out while the denominator reflects the space-time structure 
of the source and can be interpreted as its covariant homogeneity volume 
for particles of momentum $\bbox{K}$. The factor $\sqrt{\lambda}$ 
ensures \cite{M96} that only the contributions of pions from the decays 
of short-lived resonances, which happen close to the primary production 
points, are included in the average phase-space density (see section
\ref{sec2.4.1}).  


\subsection{The Usefulness of 3-Particle Correlations}
\label{sec2.6}

Two-particle correlations are insensitive to the phase of the
two-particle exchange amplitude: writing the latter for two particles
with momenta $p_{i,j}$ as
 \begin{equation}
 \label{4.18}
    \int d^4x\, S\left(x,{\textstyle{1\over 2}}(p_i + p_j)\right)\, 
    e^{i(p_i - p_j)\cdot x} = \rho_{ij}\,e^{i\phi_{ij}} \, ,
 \end{equation}
the phase $\phi_{ij}$ is seen to drop out from the correlator
(\ref{corrapp}). This is no longer true for higher-order
multiparticle correlations. For example, for a {\em completely chaotic} 
source the true 3-particle correlator, with all two-particle
correlation contributions $R_2(i,j) = C(\bbox{p}_i,\bbox{p}_j)-1$
removed, 
 \begin{equation}
 \label{R3}
    R_3(\bbox{p}_1,\bbox{p}_2,\bbox{p}_3) = 
    C_3(\bbox{p}_1,\bbox{p}_2,\bbox{p}_3) - R_2(1,2)
    - R_2(2,3) - R_2(3,1) - 1 ,
 \end{equation} 
and properly normalized, 
 \begin{equation}
 \label{r3}
    r_3(\bbox{p}_1,\bbox{p}_2,\bbox{p}_3) = 
    {R_3(\bbox{p}_1,\bbox{p}_2,\bbox{p}_3)\over
      \sqrt{ R_2(1,2)\, R_2(2,3)\, R_2(3,1)}},
 \end{equation} 
gives the sum of phases of the three two-particle exchange amplitudes
\cite{HZ97}:
 \begin{equation}
 \label{Phi}
    r_3(\bbox{p}_1,\bbox{p}_2,\bbox{p}_3) =
    2 \cos \left( \phi_{12} + \phi_{23} +\phi_{31} \right)
    \equiv 2 \cos\Phi\, .
 \end{equation} 
Expanding the two-particle exchange amplitude (\ref{4.18}) for small
relative momenta one finds \cite{HZ97}
 \begin{eqnarray}
    \Phi &=& {1\over 2}\, q_{12}^\mu\, q_{23}^\nu
                  \left[ {{\partial\langle x_\mu\rangle_3}
                          \over \partial \bar{K}^\nu}
                       - {{\partial\langle x_\nu\rangle_3}
                          \over \partial \bar{K}^\mu} \right]
 \nonumber \\
    && - {1\over 24}\, 
      \lbrack q_{12}^\mu\, q_{12}^\nu\, q_{23}^\lambda
              + q_{23}^\mu\, q_{23}^\nu\, q_{12}^\lambda\rbrack
                  \left[ {{\partial^2 \langle x_\mu\rangle_3}
                          \over \partial\bar{K}^\nu \partial\bar{K}^\lambda }
                       + {{\partial^2 \langle x_\nu\rangle_3}
                          \over \partial\bar{K}^\lambda \partial\bar{K}^\mu }
                       + {{\partial^2 \langle x_\lambda\rangle_3}
                          \over \partial\bar{K}^\mu \partial\bar{K}^\nu }
                        \right]
 \nonumber \\
    && - {1\over 2} q_{12}^\mu\, q_{23}^\nu\, 
       (q_{12} + q_{23})^\lambda\, 
       \langle \tilde{x}_\mu \tilde{x}_\nu \tilde{x}_\lambda\rangle_3 
       + O(q^4)\, .
 \label{4.24}
 \end{eqnarray}
Here $q_{ij}=p_i-p_j$ (with $q_{12}+q_{23}+q_{31} = 0$) are the
two-particle relative momenta and $\bar K = (p_1+p_2+p_3)/3$ is the
average momentum of the particle triplet; the averages
$\langle\dots\rangle_3$ are calculated with the emission function
$S(x,\bar{K})$. Eq.~(\ref{4.24}) shows that $\Phi$ depends on the odd
space-time variances $\langle \tilde{x}^3\rangle$ etc. of the emission 
function and on the derivatives with respect to $\bar{K}$ of the point
of highest emissivity ${\langle x\rangle}_3$. These reflect the
asymmetries of the source around its center. In the Gaussian
approximation of section \ref{sec2.2.1} they vanish. 

These considerations show that the true three-particle correlator 
contains additional information which is not accessible via 
two-particle correlations. In practice, however, it is difficult
to extract this information. The leading contribution to $\Phi$ is of
second order in the relative momenta $q_{ij}$, and in many reasonable 
mo\-dels it even vanishes~\cite{HV97}. Therefore new information
typically enters $r_3(\bbox{p}_1,\bbox{p}_2,\bbox{p}_3)$ at sixth 
order in $q$. The measurement of the phase $\Phi$ is thus very
sensitive to an accurate removal of all leading $q^2$-dependences by
a proper determination and normalization of the two-particle
correlator. 

On the other hand, it was pointed out that the intercept of the
normalized true three-particle correlation parameter $r_3$ may 
provide a good test for the chaoticity of the source. Writing the
emission function for a {\em partially coherent} source as $S = S_{\rm
  cha} + S_{\rm coh}$ and denoting the chaotic fraction of the
single-particle spectrum at momentum $\bbox{p}$ as $\epsilon(\bbox{p})$, 
the intercept of $r_3$ is given by \cite{HZ97,BBMST90} 
 \begin{equation}
 \label{4.25} 
    \lambda_3({\bar{\bbox{K}}}) \equiv
    r_3({\bar{\bbox{K}}},{\bar{\bbox{K}}},{\bar{\bbox{K}}})
    = 2\sqrt{\epsilon({\bar{\bbox{K}}})} 
    {{3-2\epsilon({\bar{\bbox{K}}})} \over 
     (2-\epsilon({\bar{\bbox{K}}}))^{3/2}}.
 \end{equation}
This relation is useful since, contrary to the two-particle
correlation strength $\lambda$, the intercept (\ref{4.25})
of the normalized three-particle correlator is not affected by decay
contributions from long-lived resonances which cancel in the ratio
(\ref{r3}) \cite{HZ97}. 

Complete small-$q$ expansions of $R_2$ and $R_3$ which generalize the
Gaussian parametrization (\ref{3.4}) to the case of partially coherent
sources and to three-particle correlations, improving on earlier
results in \cite{BBMST90,PRW92}, can be found in \cite{HZ97}. Within a
multidimensional analysis of 2- and 3-pion correlations they allow 
separate determination of the sizes of the homogeneity regions of the
chaotic and coherent source components as well as the distance between
their centers.  


\section{TWO-PARTICLE CORRELATIONS FROM DYNAMICAL MODELS}
\label{sec3}

Interpretation of correlation functions measured in heavy-ion
collisions requires understanding the true relationship of the 
parameters extracted from fitting the data and the actual
single-particle distributions at freeze-out. The level to which this 
works in practice can be established by using an event generator 
to model the collision dynamics, particle production and hadronic 
rescattering, and then constructing a two-particle correlation function. 
These functions can be fit in the same way as experimental correlation 
functions and the fit parameters compared to the single-particle 
freeze-out distribution in the event generator \cite{sulli,zajc}. 

The event generator correlation functions are constructed from the
positions and momenta representing the single particle emission
distribution at the time of the last strong interaction ({\it i.e.} at
freeze-out). The subsequent calculation of correlation functions uses
particle pairs drawn randomly from this list and constructs a two-particle 
symmetrized wave function \cite{spacer,PCZ90,sulli,WH99}. Coulomb
wave functions for the particles are used. As for experimental data, 
a Coulomb correction is applied to the correlation function before 
fitting \cite{sulli}.

As the event generator yields a correlation function while
simultaneously knowing the space-time distribution and history of the
same particles, discrepancies between the fit parameters and
freeze-out distributions may be resolved. Furthermore, the generated
particles may be subjected to experimental acceptance cuts and treated 
like real data. This allows evaluation of the effects of experimental 
acceptances and analysis techniques. Since a significant number of 
the observed hadrons arise from decay of long-lived resonances, the event
generators can also quantify their effects on the correlation
functions. Such studies were performed using the RQMD 
\cite{rqmd,sulli,na35l,na44mt,na44pb} and ARC event generators 
\cite{arc,zajc} as well as hydrodynamical simulations 
\cite{schlei,RG96,SOPSW96,OPSSW96,SX96,schleiEOS}.
 
\subsection {RQMD}
\label{sec3.1}

Many experiments use the RQMD event generator \cite{rqmd}, as it
satisfactorily reproduces single-particle distributions. RQMD simulates 
the microscopic phase-space evolution, using resonance and string 
excitation as primary processes, followed by fragmentation, decays and 
subsequent hadronic collisions. Many features in p-nucleus and 
nucleus-nucleus collisions which can be related to secondary scattering 
are well described by RQMD \cite{mattiel,sorgelett}. The numerous 
secondary collisions result in considerable transverse flow of RQMD 
events before freeze-out \cite{sulli,814l,fields}. 

\subsubsection {Collective expansion}
\label{sec3.1.1}

\begin{figure}[ht]
\vspace*{8cm}
\includegraphics{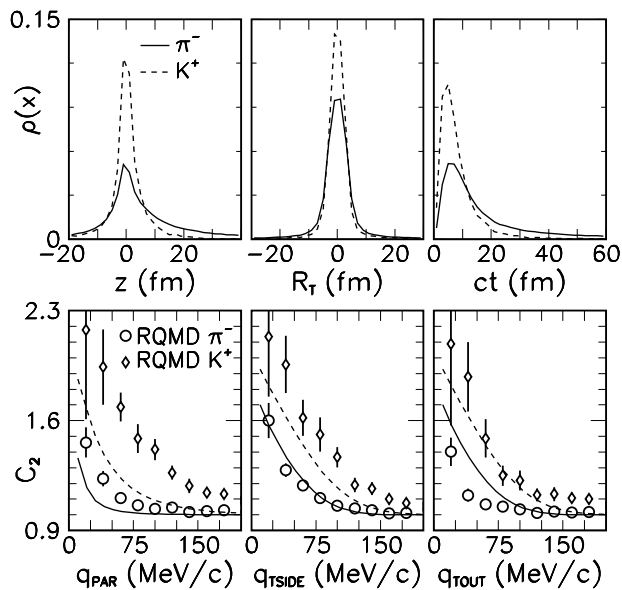}
\caption{Comparison of position distributions at freeze-out and 
         calculated correlation functions, both from RQMD events.
         The lines in the bottom figures show $1+\vert\tilde\rho(q)\vert^2$
         where $\tilde\rho(q)$ is the Fourier transform of the 
         corresponding distribution in the upper figures. The notation on 
         the axes $q_{_{\rm PAR}}, q_{_{\rm TSIDE}}, q_{_{\rm TOUT}}$ 
         corresponds to $q_l, q_s, q_o$ used elsewhere in this
         review. (Figure taken from \protect\cite{sulli}.)
\label{rqmd}}
\end{figure}

Figure~\ref{rqmd} shows a comparison of RQMD freezeout positions and
correlation functions for S+Pb collisions at 200 GeV/nucleon
\cite{sulli}. The top half shows the pion (solid lines) and kaon
%
%
(dotted lines) position distributions at freeze-out along and transverse
to the beam direction as well as in time, integrated over their
momenta. The kaon distributions are narrower than those of the pions. 
The lower section of the figure shows correlation functions calculated 
from the RQMD events, plotted as functions of the Cartesian variables. 
The points show the calculated correlation functions using the NA35 
experimental acceptance, while the solid and dashed lines indicate 
Fourier transforms of the relevant components of the pion and kaon 
freeze-out distributions in the top row of figures. If the particle 
positions and momenta were uncorrelated, HBT interferometry should
reproduce the full size of the freeze-out distribution, and in the 
lower panels the lines should agree with the corresponding points.
Instead, the correlation functions calculated from the RQMD 
{\em phase-space} distribution are much wider in $q_l$ than 
expected from the Fourier transform, indicating a smaller effective 
source. This reflects longitudinal position-momentum correlations 
arising from a strong longitudinal expansion of the source. As will be 
seen later, this prediction by RQMD and hydrodynamical 
models is confirmed by the data. An analogous effect is seen
in the sideward direction where especially the calculated kaon 
correlation function is wider than expected from the Fourier transform 
of their momentum-integrated transverse spatial distribution. This can
be traced back to transverse position-momentum correlations in the
RQMD freeze-out distribution, induced by collective transverse expansion.
As discussed above, these correlations increase with increasing
transverse mass of the particles and are thus more strongly reflected
in the kaon correlation function. The resulting decrease of the 
transverse effective source size extracted from kaon correlations
is again confirmed by the data. 

Event generators like RQMD can be used to study the influence of 
experimental acceptance cuts on the $K_\perp$-dependence of the
HBT fit parameters induced by collective flow. RQMD was shown 
to reproduce the $K_\perp$-dependence of the correlation functions 
both at SPS \cite{na35l,na44mt,na44pb} and AGS \cite{baker,MLisa} 
energies. It is noteworthy that by analyzing the {\em same} 
set of RQMD events with the NA35 and NA44 acceptances, good agreement 
with both data sets was obtained even though the $K_\perp$-dependence 
of $R_s$ appears to be somewhat stronger in NA44 \cite{fields}. 

\begin{figure}[ht]
\centerline{\epsfxsize=11cm\epsffile{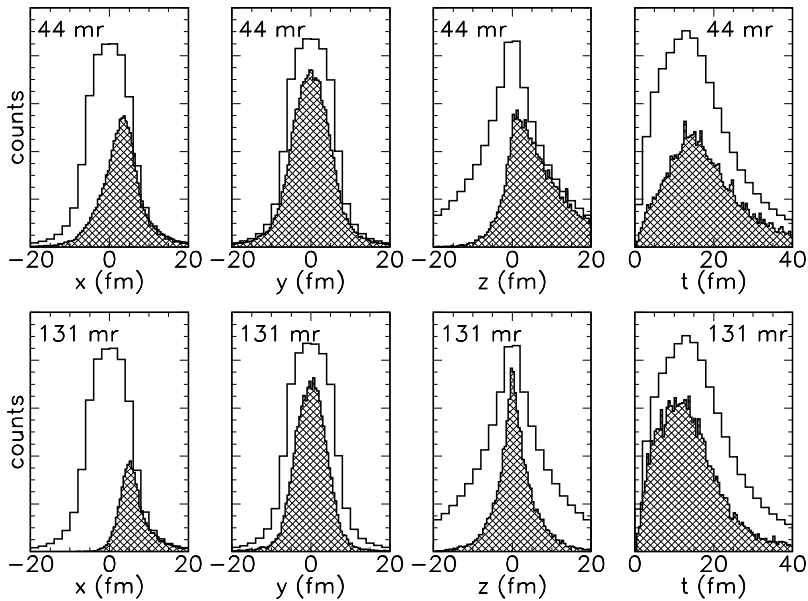}}
\caption{RQMD freeze-out distributions for pions. The open histograms 
   give all pions from RQMD, the hatched ones only those accepted by
   NA44 at various spectrometer settings. $x,y,z$ indicate the
   out-, side- and longitudinal directions in the collision center 
   of mass system. (Figure taken from \protect\cite{na44pb}.)
\label{freeze}}
\end{figure}

The shrinking of the effective source size as a result of collective 
flow is illustrated in Figure~\ref{freeze}. The collective flow velocity 
has the effect of ``focussing'' particles arising from nearby regions 
of the source. As the correlation selects particles of small momentum 
difference, it is sensitive to this focussing. 
%
%
Figure~\ref{freeze} shows the freeze-out distributions of all 
pions (open histograms) and of those pions accepted in two NA44 
spectrometer settings (44mr is the low $K_\perp$ setting and 131mr 
high $K_\perp$). Higher $K_\perp$ pairs have a narrower distribution 
at freeze-out due to the larger effect of the flow velocity. Note
that this is the focussing from flow and not an experimental 
acceptance effect. The correlation function only ``sees'' particles
which are flowing in the same direction; particles from the far 
side of the source flow away in an expanding source. This is why the
characteristic drop of apparent source size with increasing $K_\perp$
is observed by both NA35 and NA44 even though one experiment has large
acceptance and one narrow. 

\subsubsection {Emission duration}
\label{sec3.1.2}

The emission duration is usually calculated from the fit
parameters via Eq.~(\ref{Rdiff}), using the leading term
only. Comparison of this emission time estimate with the actual width
of the freeze-out time distribution from RQMD events determines the
validity of neglecting the position-momentum correlation effects on
the space-time variances. Fields et al. \cite{fields} found that only
for $K_\perp$ below $\approx$ 100 MeV/c does this method yield the
actual emission duration. For higher $K_\perp$ particles, the true
emission duration (3-6 fm/$c$ for S+Pb \cite{fields} and 7-8.5 
fm/$c$ for Pb+Pb \cite{na44pb}) is shorter due to flow-induced 
position-momentum correlations, but the value derived from the 
difference between $R_o^2$ and $R_s^2$ underrepresents even this. 
Consequently, extraction of the emission duration from experimental 
correlation functions should only be attempted at low $K_\perp$.  

Of course, the total lifetime of the source between impact and freeze-out 
is considerably larger.


\subsubsection {Resonance decays}
\label{sec3.1.3}

In RQMD as many as half of the low $K_\perp$ pions in heavy-ion 
collisions arise from the decay of long-lived resonances \cite{sulli}. 
Very long-lived resonances produce a long tail in 
the pion freeze-out position distribution, visible in Figure~\ref{rqmd}. 
This corresponds to a component of the correlation function too narrow 
to measure and reduces the correlation strength $\lambda$ \cite{sulli}. 
In S+Pb collisions the fraction of pions from $\omega, \eta, \eta'$ 
resonances is 30\% at low $p_\perp$ and falls to 5 \% at 
$p_\perp$= 800 MeV/c \cite{fields}. These pions cause a departure 
from a Gaussian source shape, which impacts the quality of
fits with a Gaussian parametrization.

Resonance decay contributions to kaons are smaller, which makes
the interpretation of their correlation functions cleaner. Due to 
their rest mass they always have a sizeable $M_\perp$, and so their 
correlation functions are more sensitive to flow.


\subsection{Hydrodynamical Models}
\label{sec3.2}

A more macroscopic approach to describe the dynamical evolution of the
reaction zone treats it as a locally thermalized ideal fluid and solves 
the relativistic hydrodynamical equations. Initial conditions are usually
set after an initial pre-equilibrium stage and suitably parametrized 
\cite{schlei,SOPSW96,RG96,Setal97}. Freeze-out is usually enforced at a 
fixed energy density or temperature. After the end of the simulation
those fluid cells which satisfy this freeze-out criterium are identified, 
and the local energy and baryon density in these cells are converted
into temperature and chemical potentials using the equation of state   
of an ideal resonance gas in thermal and chemical equilibrium. Each such
cell thus emits a thermal hadron spectrum boosted by the local fluid 
velocity. This determines the emission function $S(x,p)$ of the model 
from which spectra and correlation functions can be calculated 
\cite{schlei,SOPSW96,OPSSW96,SX96}.

Measured spectra and correlation functions put constraints on the 
output of such simulations which can be used to identify allowed 
combinations of initial conditions and equations of state 
\cite{Setal97,schleiEOS}. In this way certain classes of evolution 
scenarios can be eliminated while successful combinations can be 
used to predict other observables for further discrimination or
hadronic one- and two-particle spectra at future colliders \cite{SS99}.

\section{TWO-PARTICLE CORRELATIONS IN HEAVY-ION EX\-PE\-RIMENTS}
\label{sec4}

\subsection{General Remarks and Short Overview of the Experiments}
\label{sec4.1}
\subsubsection{Construction of the correlation function}
\label{sec4.1.1}

Experimentally, correlation functions are constructed by counting
events with boson pairs of given pair and relative momenta and
dividing by a properly normalized \cite{MV97,Z84} ``background'' 
sample with no enhancement: 
 \begin{equation}
   C_2^{\rm expmt.}(\bbox{q},\bbox{K}) = 
   A(\bbox{q},\bbox{K})/B(\bbox{q},\bbox{K}).
 \label{eq:4}
 \end{equation}
Typically, $B(\bbox{q},\bbox{K})$ is generated by creating artificial
pairs by combining single tracks from different events, or by using
unlike-sign pions. Generally, analyses bin both the data and the
background in the chosen variables. The correlation functions are
then corrected for Coulomb interactions, experimental resolution and
two-particle acceptance (generally by Monte Carlo techniques), 
residual two-particle effects on the single-particle spectrum 
constructed from mixed pairs, particle misidentification (if
contamination is significant); see \protect\cite{e802pi,NA441d,na44k,
na35zeit} for details.

\subsubsection{Fit procedures}
\label{sec4.1.2}

Correlation functions are customarily fit by a Gaussian in $q$ which 
assumes a Gaussian-distributed source. This is a simplification, since 
the source may well have a more complex shape. However, the Gaussian
assumption provides a reasonable representation of the data 
\cite{NA441d,NA49HBT} and continues to be used. One should keep in mind, 
however, that the measured correlation functions are usually not perfect 
Gaussians and that this leads to systematic uncertainties of the order 
of 10-20\% in the extracted fit parameters.

High statistics data samples are now analyzed using multi-dimensional 
fits\cite{P84,CSH95b,CNH95,HTWW96,ykp}. Unlike
many lower-dimensional parametrizations these do not require the 
unrealistic assumption of a spherical source and are more sensitive 
to the collision dynamics \cite{zajc}. Some analyses do, however, 
use lower-dimensional parameterizations due to statistical or 
acceptance limitations. These carry the danger of producing 
misleading results, but they can be useful if they are based on a
complete parametrization and make proper projections \cite{baker}.

\subsubsection{Coulomb corrections in experiment} 
\label{sec4.1.3}

Initially, experimental correlation functions were corrected for 2-body
Coulomb interactions using the Gamov factor. However, as measurements 
of heavier systems were made, the point source approximation became 
increasingly inappropriate. Several different techniques have been 
used for improved corrections.

The first experimental improvement was achieved by integrating 
expression (\ref{Koonin}) with Coulomb wave functions using a 
technique developed by Pratt \cite{coulwave}. For the relative 
source function $S_{\bbox{K}}(\bbox{r})$ one usually takes a 
spherically symmetric Gaussian in the pair rest frame with a size 
parameter which is iterated. In S+Pb collisions, NA44 found that 
the improved Coulomb correction resulted in a 5-10\% increase in 
the fitted radius parameters for pions and kaons 
\cite{NA441d,na44k,na44mt}. In Pb+Pb collisions, the difference is
8-12\%, with $\lambda$ decreasing by 3-6\% \cite{na44pb}.

The Coulomb correction can be investigated experimentally by measuring
correlation functions of oppositely charged particle pairs, where quantum
mechanical symmetrization effects are absent \cite{NA35newcoul}. The 
measured Coulomb attraction reflects the (non-zero) source size 
\cite{BBM96}, and can be used to parametrize a Coulomb correction 
for same sign pairs. This technique is used by NA49 \cite{NA49HBT}.


\subsubsection{Short description of the experiments}
\label{sec4.1.4}

We now briefly describe the experiments whose data will be discussed
in the following. E802/859/866 at the AGS is a wide acceptance magnetic 
spectrometer experiment, tracking particles with drift chambers and 
identifying them via time-of-flight. E814/877 at the AGS
is a multipurpose experiment which includes a hadron spectrometer
covering forward angles. Particle identification is achieved using
time-of-flight with a very long flight path. The angular range of the
acceptance limits the identified hadrons to rather small $p_\perp$. 
E895 is a time projection chamber (TPC) at the AGS, which identifies
hadrons through their energy loss in the gas-filled detector volume.

At the CERN SPS, NA44 is a second generation experiment which measures
single and two-particle distributions at midrapidity. It is
characterized by excellent particle identification, with
contaminations at the 1\% level. As a focussing spectrometer, its
acceptance for particle pairs with small momentum difference is
optimized, allowing for high statistics in the region of the
Bose-Einstein correlation signal. NA35 is a streamer chamber and TPC
experiment at CERN, with a large acceptance for pions from 200~$A$ GeV
S-nucleus collisions. NA49, the successor to NA35, uses four TPC's and 
two time-of-flight walls to track and identify particles in Pb+Pb
collisions. For the correlation functions NA35 and NA49 use charged 
tracks without particle identification; they are fit as a function 
of pair rapidity and transverse momentum.   

Different experiments analyze correlations in different reference
frames. The NA44 Collaboration \cite{na44k,na44mt} use the LCMS in
which the longitudinal pair momentum vanishes. This frame couples the
lifetime information solely to $q_o$ and ensures that the source
velocity in the analysis frame is usually small \cite{ykp}. The
NA35/NA49 Collaboration \cite{na35zeit,na35l} and E802/E859/E866
\cite{e802pi,e802k,e859qm} analyze in the nucleon-nucleon center of 
mass frame which is similar (but for asymmetric collision systems not
identical) to the LCMS at mid-rapidity.

\subsubsection{Square roots of 2, 3, and 5}
\label{sec4.1.5}

Confusion can easily arise when comparing fitted HBT radius parameters
with the rms or hard sphere radii of the colliding nuclei. As is well
known, the 3-dimensional rms radius $R_{\rm rms,3d}$ and the hard
sphere radius $R_{\rm box}=1.2\,A^{1/3}$~fm are related by a factor
$\sqrt{3/5}$:  
 \begin{equation}
 \label{eq3.1}
   R_{\rm rms,3d}^2 = \langle \bbox{r}^2\rangle 
   = \langle x_o^2+x_s^2+x_l^2 \rangle
   = {\int_0^{R_{\rm box}} r^2\, d^3r\over \int_0^{R_{\rm box}} d^3r}
   = {3\over 5} R_{\rm box}^2.
 \end{equation}
The HBT radii are {\em 1-dimensional rms radii} (e.g. $R_s^2=\langle
x_s^2 \rangle$) and thus another factor $\sqrt{3}$ smaller than $R_{\rm
  rms,3d}$. If a cold spherical nucleus in its ground state could be 
induced to emit pion pairs, one would thus measure
 \begin{equation}
 \label{eq3.2}
   R_s^{\rm cold} = {R_{\rm rms,3d}\over \sqrt{3}}
   = {R_{\rm box}\over \sqrt{5}}.
 \end{equation}
We will call the ratio of the actually measured $R_s$ to this naive 
expectation the ``expansion factor'' $\xi$.

At high energies it is often useful to compare with the
2-dimensional rms radius of the nuclear overlap region in the
transverse plane, $R_{\rm rms,2d} = \langle x_o^2+x_s^2\rangle^{1/2}$,
which is a factor $\sqrt{2}$ larger than the corresponding sideward
radius $R_s=R_\perp$.


\subsection{A Measured Correlation Function}
\label{sec4.2}

Collision systems with light projectiles (S+S and S+Pb at the SPS and 
Si+Au at the AGS) have been studied by several experiments, with 
some systematic differences in the results. NA35 included a factor of 
${1\over 2}$ in the exponent of the Gaussian fit function, which 
yields $R$ parameters larger by $\sqrt{2}$ than those from other 
experiments. 

\begin{figure}[ht]
\vspace*{9cm}
\includegraphics{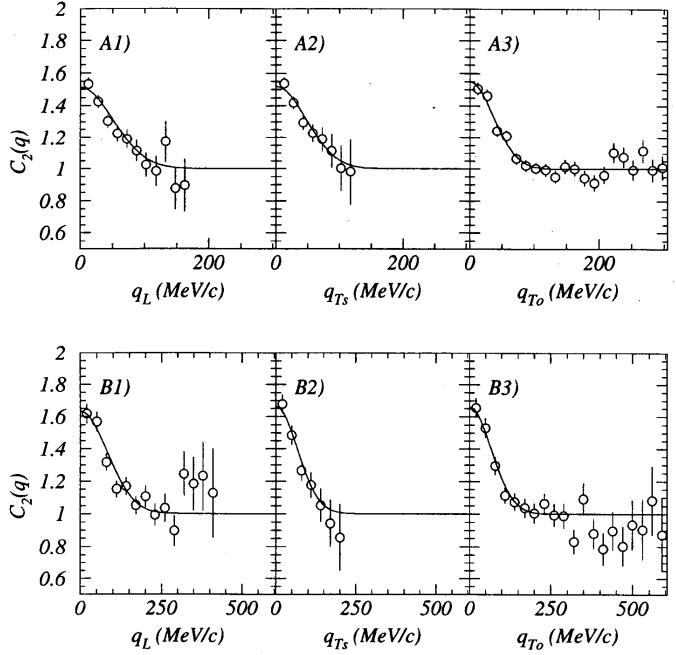}
\caption{$\pi^+\pi^+$ and $K^+K^+$ correlation functions for 14.6 $A$ 
  GeV Si+Au collisions from E859 \cite{e859qm}. The analysis was done 
  in the participant center of mass ($Y$=1.25). For the projections the 
  remaining two components of the relative momentum were cut to 
  $5$ MeV$/c<q_{\rm perp} < 35\ (65)$ MeV/$c$ for pions (kaons). The solid 
  lines show a 3-d Gaussian fit. (Figure taken with permission from 
  \protect\cite{e859qm}.)
\label{F4}}
\end{figure} 

Figure \ref{F4} shows three-dimensional correlation functions of 
$\pi^+$ and $K^+$ pairs in 14.6 $A$ GeV Si+Au collisions, measured by 
E859 \cite{e859qm}. Shown are the projections in the three Cartesian 
directions, with narrow cuts on the momentum differences in the 
other two directions. However, the fits are performed in three 
dimensions, not upon the projections. The solid line shows a
Gaussian fit in the CM system ($Y$=1.25) according to (\ref{modCart}) 
without the cross-terms; the shape of the data is reproduced quite well. 

\begin{table}[ht]
 \begin{center}\mbox{}
  \begin{tabular}{|c|c|c|c|c|}\hline
pair & $R_l$ & $R_s$ & $R_o$ & $\lambda$ \\ \hline
$\pi^+\pi^+$ &2.75$\pm$0.15 &2.95$\pm$0.19 &3.77$\pm$0.13 &0.65$\pm$0.02\\
$K^+K^+$ & 1.71$\pm$0.14 & 2.09$\pm$0.20 &2.07$\pm$0.16 &0.83$\pm$0.08\\
\hline
  \end{tabular}
 \end{center}
 \caption{E859 correlation function fit parameters \protect\cite{e859qm}.
          Pions have rapidities $1.2<y<1.8$ and transverse momenta
          $100$ MeV$/c < p_\perp < 800$ MeV/$c$, while kaons are 
          accepted for $1.0<y<1.7$ and $100$ MeV$/c<p_\perp <900$ MeV/$c$.
 \label{T1}}
\end{table}

The fit parameters extracted by E859 from pion and kaon correlations 
are shown in Table~\ref{T1}. The radius parameters from kaons are 
considerably smaller than those from pions, but the $\lambda$ parameter 
is larger \cite{e802pi,e802k}. Both trends are expected from the fact 
that kaons are less affected by resonance decays than pions 
\cite{Gyulassy&Padula}; from the discussion in Sec.~\ref{sec2.4.1}, 
however, the effect on the radius parameters should have been weaker. 
It was also postulated that, due to their smaller cross section with 
nucleons, kaons may freeze out earlier than pions \cite{Shoji}, reflecting 
a smaller source if the latter expands. The E859 analysis \cite{e859qm} 
showed, however, within a fit of reduced dimensionality (2 instead of 
3 radius parameters) that also the pion correlation radii decrease 
systematically with increasing transverse mass of the pair. However, 
the interpretation of such a 2-dimensional fit is not straightforward.


\subsection{Asymmetric Collision Systems: First Signs of 
            $M_\perp$-Dependence} 
\label{sec4.3}

The first 3-dimensional and $\bbox{K}$-dependent correlation analysis 
was achieved in sulphur-in\-duced collisions at the SPS. Table~\ref{T2} 
summarizes the $R$ parameters extracted by NA44 from a 3-dimensional 
analysis of pion and kaon correlations in 450 GeV p+Pb and 200 $A$ GeV 
S+Pb collisions at the SPS \cite{na44pi,na44k}. They were extracted from 
a fit to 
 \begin{equation}
 \label{3d}
   C(\bbox{q},\bbox{K}) = 1 + \lambda(\bbox{K})
   \exp(-R_s^2(\bbox{K})q_s^2 - R_o^2(\bbox{K})q_o^2
        -R_l^2(\bbox{K})q_l^2)\,.
 \end{equation}
One sees the same trends as observed by E802/859 at lower beam energy 
\cite{e802k,e859qm}:
%
\begin{table}[ht]
 \begin{center}\mbox{}
  \begin{tabular}{|c|c|c|c|c|c|}\hline
        & & $R_s$   & $R_o$     & $R_l$ &$\lambda$\\ \hline
S + Pb &$\pi^+$&      4.15$\pm$0.27&  4.02$\pm$0.14&  4.73$\pm$0.26&
        0.56$\pm$0.02\\
p + Pb &$\pi^+$&      2.00$\pm$0.25&  1.92$\pm$0.13&  2.34$\pm$0.36&
        0.41$\pm$0.02\\
S + Pb &$K^+$&  2.55$\pm$0.20&  2.77$\pm$0.12&  3.02$\pm$0.20&  
   0.82$\pm$0.04\\
p + Pb &$K^+$& 1.22$\pm$0.76&   1.53$\pm$0.17&  2.40$\pm$0.30&  
   0.70$\pm$0.07\\
\hline
  \end{tabular}
 \end{center}
 \caption {HBT parameters in the LCMS measured by NA44 for 
           450 GeV p+Pb and 200 $A$ GeV S+Pb collisions 
           \protect\cite{na44pi,na44k}. Pions are measured in 
           $3.2<y<4.2$, kaons in $2.7<y<3.3$. The $p_\perp$ acceptance 
           is $0<p_\perp <0.6$ GeV/$c$ for both.
 \label{T2}}
\end{table}
%
the $R$ parameters for kaons are consistently smaller than for pions 
and the correlation strength $\lambda$ is larger. That the $R$ parameters 
are larger in S+Pb than in p+Pb collisions should be expected. However,
$R_s$ in S+Pb collisions is also (much) larger than the projectile, even 
for kaons: for $^{32}$S we should compare $R_s$ to $R_{\rm box}/\sqrt{5}
=3.8\,{\rm fm}/2.23 = 1.7$ fm. Thus the system must have expanded 
significantly before freezeout. 

\begin{figure}[ht]
\vspace*{8cm}
\includegraphics{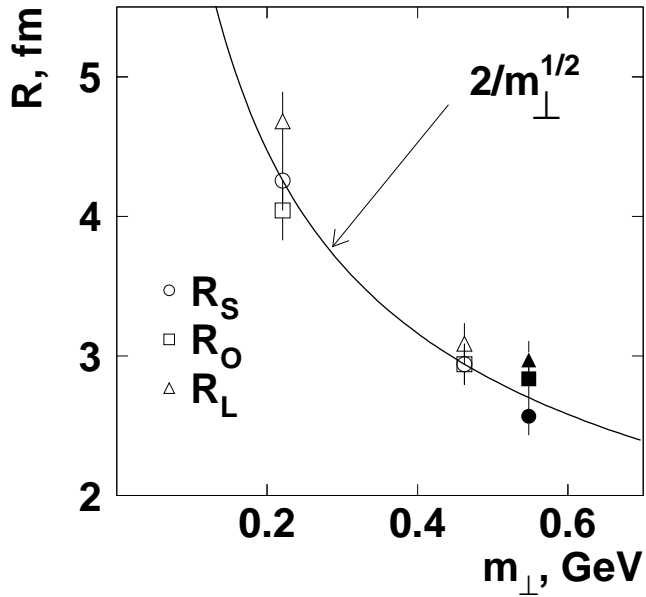}
\caption{$M_\perp$-dependence of the three Cartesian radius parameters
    in 200 $A$ GeV S+Pb collisions measured by NA44 \cite{na44mt}. The
    open (filled) symbols indicate values extracted from $\pi^+\pi^+$ 
    ($K^+K^+$) correlations. The solid line is given by 
    $R = 2$ fm/$\sqrt{M_\perp}$ ($M_\perp$ in GeV).
\label{F5}}
\end{figure}

This is supported by the additional observation (see Figure~\ref{F5}) that
%
%
all three $R$ parameters show a strong dependence on the transverse 
mass $M_\perp$ of the pairs. In \cite{na44mt} they were compared to a common 
$1/\sqrt{M_\perp}$-law (see Figure~\ref{F5}) which simultaneously 
reproduces the two pion points and one kaon point in each of the three 
Cartesian directions. This points to a dependence on $M_\perp$ (rather 
than on $m_0$ and $K_\perp$ separately), as expected for a thermalized 
expanding source \cite{CL95}. 

NA44 compared correlations of $K^+$ and $K^-$ pairs, to 
investigate possible effects of their different cross sections with 
nucleons on the freeze-out distribution. No significant 
differences were found, indicating that at SPS energies the $KN$ cross 
section plays a subdominant role in the freeze-out process. This is 
to be expected from the small measured nucleon/pion ratio: $p/\pi^+ = 0.12$ 
\cite{na44prot}; $K\pi$ scattering dominates and is similar for $K^+$ 
and $K^-$. 

The fit function (\ref{3d}) did not include the cross-term $R_{ol}$.
It is expected to be small near mid-rapidity where the NA44 acceptance 
is concentrated. Subsequent analyses showed \cite{na44pb} that including 
the cross-term changes the $R$ parameters by $\lapp10\%$, less than the 
20\% systematic uncertainty on their absolute values.  

NA35 has measured the rapidity and transverse momentum dependence of 
the three source radius parameters $R_s$, $R_o$, and $R_l$ for central 
S+S, S+Cu, S+Ag, and S+Au collisions\cite{na35zeit}, by fitting the 
correlation functions to Eq.~(\ref{3d}) in the nucleon-nucleon center
of mass. The rapidity dependence of the transverse radius parameters is 
minimal. $R_l$ depends strongly on the pair rapidity and is well described
by $R_l \sim 1/\cosh(Y{-}Y_{_{\rm CM}})$, indicating approximately
boost-invariant longitudinal expansion \cite{MS88} as was predicted 
for collisions at these energies \cite{bj}. If this $Y$-dependence in 
the CM frame were exact, $R_l$ would be independent of $Y$ in the LCMS.
One should remember, however, that the analysis of \cite{na35zeit} was 
done without including the $R_{ol}$ cross-term which was shown 
\cite{AlberQM95} to become sizeable in the CM at large $Y-Y_{_{\rm CM}}$.
 
All systems show a similar dependence of $R_l$ on the transverse pair 
momentum $K_\perp$, with $R_l$ decreasing with increasing $K_\perp$ 
at a similar rate as found by NA44 (see above). This provides 
further evidence for longitudinal expansion of the source (see 
Sec.~\ref{sec2.3}), and this behaviour is also predicted by
hydrodynamical simulations of the collisions\cite{OPSSW96,na35zeit}. 

Compared to $R_l$, the transverse parameters $R_s$, $R_o$ from the 
NA35 analysis \cite{na35zeit} show a weaker $K_\perp$-dependence, 
indicating that the longitudal and transverse expansions differ. 
Such a tendency is generically expected from hydrodynamic 
simulations of the collision dynamics \cite{OPSSW96,SX96} and from 
the model studies presented in Sec.~\ref{sec2.3}. As discussed in 
section \ref{sec3.1.1}, the apparently stronger $K_\perp$-dependence 
of the transverse $R$ parameters observed by NA44 in S+Pb collisions 
\cite{na44mt} can be understood in terms of the different experimental 
acceptances.


\subsection{Au+Au Collisions at the AGS}
\label{sec4.4}

Collective transverse expansion of the emitting source was also observed 
at the AGS. A characteristic $K_\perp$ dependence of the transverse radius
parameter $R_\perp$ was found in Au+Au collisions at 11 GeV/nucleon by 
E866 \cite{baker}. Decreasing values of all three Cartesian source 
parameters with increasing $K_\perp$ are observed in Au + Au collisions 
at energies as low as 2 GeV/nucleon by the E895 collaboration \cite{MLisa},
although at their lowest beam energy of 2 Gev/nucleon the $K_\perp$
dependence of $R_s$ and $R_o$ seems to disappear. Though the hadron 
densities at such energies are much less than at 158 GeV/nucleon, they 
are still quite large. Even here, the hadronic scattering can generate 
pressure and cause the source to expand. In fact, pair 
momentum dependence of the fit parameters was already observed in 
streamer chamber data at the Bevalac by Beavis et al. \cite{sc} in a 
1-dimensional analysis of pion correlations from 1.8 GeV/nucleon Ar+Pb
collisions. Already at that time the dependence was interpreted as 
evidence for expansion of the source before freeze-out \cite{sc}.


\subsection{Pb+Pb Collisions at the SPS}
\label{sec4.5}
\subsubsection{Cartesian parametrization}
\label{sec4.5.1}

\begin{figure}[ht]
\centerline{\epsfxsize=13cm\epsffile{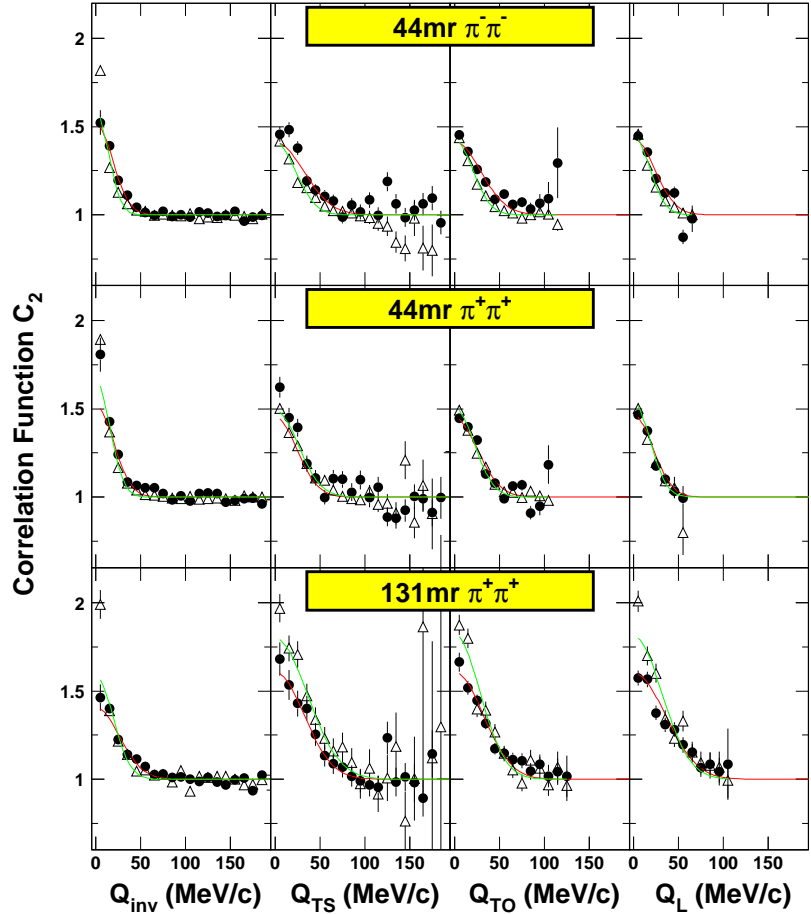}}
\caption{Comparison of NA44 Pb+Pb $\pi^+\pi^+$ correlation data (solid 
   circles) and RQMD predictions (open triangles). The 1-dimensional 
   projections of the 3-dimensional correlation function are averaged 
   over the lowest 20 MeV/$c$ in the other momentum differences. (Figure
   taken from \cite{na44pb}.)  
\label{na44pb}}
\end{figure}

Figure~\ref{na44pb} shows projections of the 3-dimensional correlation 
functions for $\pi^-$ and $\pi^+$ pairs from 158 GeV/nucleon Pb+Pb 
collisions, measured by NA44 (closed points) \cite{na44pb}. 
%
%
The left column shows a 1-dimensional analysis in $q_{\rm inv}$ for
comparison with older data from other colliding systems. Two angular 
settings of the spectrometer cover $p_\perp$-ranges of 0-0.4 GeV/c 
and 0.3-0.8 GeV/c. The fitted source parameters, with and without the 
$R_{ol}$ cross-term, are given in Table~\ref{T3} \cite{na44pb}.
%
\begin{table}[ht]
 \begin{center}\mbox{}
  \begin{tabular}{|c|c|c|c|c|c|}\hline
 pair\ ($\langle K_\perp\rangle$) & $\lambda$ & $R_o$ (fm) & $R_s$ (fm) 
     & $R_l$ (fm) &$R^2_{ol}$ (fm$^2$)\\ \hline
$\pi^-\pi^- (\approx 170)$ &0.495$\pm$0.023 &4.88$\pm$0.21 & 
4.45$\pm$0.32 &6.03$\pm$0.35 &\\
$\pi^+\pi^+ (\approx 170)$ &0.569$\pm$0.035 &5.50$\pm$0.26 & 
5.87$\pm$0.58 &6.58$\pm$0.48 &\\
$\pi^+\pi^+ (\approx 480)$ &0.679$\pm$0.034 &4.39$\pm$0.18 & 
4.39$\pm$0.31 &3.96$\pm$0.23 &\\
$\pi^-\pi^- (\approx 170)$ &0.524$\pm$0.026 &5.35$\pm$0.25 & 
5.07$\pm$0.35 &6.68$\pm$0.39 &10.7$\pm$2.9\\
$\pi^+\pi^+ (\approx 170)$ &0.658$\pm$0.035 &5.98$\pm$0.23 & 
6.94$\pm$0.48 &7.39$\pm$0.40 &28.1$\pm$3.5\\
$\pi^+\pi^+ (\approx 480)$ &0.693$\pm$0.037 &4.59$\pm$0.21 & 
4.71$\pm$0.36 &4.15$\pm$0.25 &3.1$\pm$1.4\\
\hline
  \end{tabular}
 \end{center}
 \caption{Cartesian radius parameters from a Gaussian fit to NA44 
    correlation functions for Pb+Pb collisions \protect\cite{na44pb} 
    using the Coulomb wave correction. The fitted results with and 
    without the cross-term $R_{ol}^2$ are shown. 
    ($\langle K_\perp \rangle$ in MeV/$c$.)
\label{T3}
}
\end{table}
%
As in the case of sulphur-induced collisions, all $R$ parameters are 
seen to become significantly smaller as $K_\perp$ increases, again 
pointing to collective longitudinal and transverse flow of the source. 
At small $K_\perp$ $R_s$ is once again much larger than the corresponding 
value (3.1 fm) of a cold Pb nucleus, indicating that the transverse 
{\em flow} also leads to transverse {\em growth} of the collision zone 
before freeze-out.

The cross-term $R^2_{ol}$ is non-zero for all data sets, and it is 
rather large for the low-$K_\perp$ $\pi^+$ pairs. As explained above, 
this term should be non-zero in the LCMS frame (where these data
were analyzed) if the source is not reflection symmetric in beam 
direction. Since the NA44 low-$K_\perp$ acceptance is slightly 
forward of midrapidity, this condition of symmetry is not fulfilled. 
The fitted $R$ parameters all become larger when the cross-term is 
included. The higher $K_\perp$ acceptance is nearer midrapidity, and 
the cross-term is indeed smaller.

It should be noted that the fit parameters for positively and negatively
charged pions differ in an apparently significant way. However, 
calculation of the $\chi^2$ per degree of freedom between the two 
{\em measured} correlation functions yields a value of 450/440 
\cite{na44pb}. As this is near unity, the experimenters concluded that 
the correlation functions do not, in fact, differ. This illustrates an 
important systematic limitation in extracting source parameters from 
Gaussian fits to measured correlation functions. Such problems are 
certainly exacerbated when comparing data from different experiments 
where statistical and systematic errors depend differently on $\bbox{q}$. 
These limitations hold regardless of the choice of source 
parametrization, but a comparison of different parametrizations 
(in which the available $\bbox{q}$-space is differently populated)
via the cross-check relations given in Sec.~\ref{sec2.2} could 
provide an estimate for the corresponding systematic uncertainties. 

Figure~\ref{na44pb} also shows pion correlation functions calculated 
from the RQMD event generator \cite{rqmd}, using the same charged 
multiplicity as selected by the experiment and a filter simulating the 
NA44 acceptance. RQMD predicts source size parameters which are 
slightly larger than the measured ones \cite{na44pb}, but agrees 
remarkably well with the general trend of the data. It reproduces
the larger values of $R_l$ compared to the transverse $R$ parameters 
for low $K_\perp$ and predicts radius parameters similar to the measured 
ones at high $K_\perp$. It overpredicts, however, significantly the 
value of the correlation strength $\lambda$. This discrepancy is 
likely due to non-Gaussian distortions of the correlation function
by pions from resonance decays and the resulting systematic uncertainties 
in Gaussian fits (see Sec.~\ref{sec2.4.1}).


\subsubsection{Yano-Koonin-Podgoretski\u{\i}\  parametrization}
\label{sec4.5.2}

An analysis of charged particle correlations from 158 $A$ GeV Pb+Pb 
collisions with the Yano-Koonin-Podgoretski\u{\i}\ parametrization 
(\ref{YKP}) (amended by a correlation strength parameter $\lambda$) 
was performed by the NA49 Collaboration \cite{NA49QM96,NA49HBT}. The 
analysis is not based on identified pions, and the non-pion contamination 
(which contributes to the mixed-pair background but not to the 
correlated pairs) reduces the value of $\lambda$ significantly. However, 
the values of the YKP fit parameters are affected only at the 
2-6\% level \cite{NA49HBT}.

\begin{figure}[ht]
\begin{center}
   \begin{minipage}[t]{7.0truecm}
	 \epsfxsize 7.0truecm \epsfbox{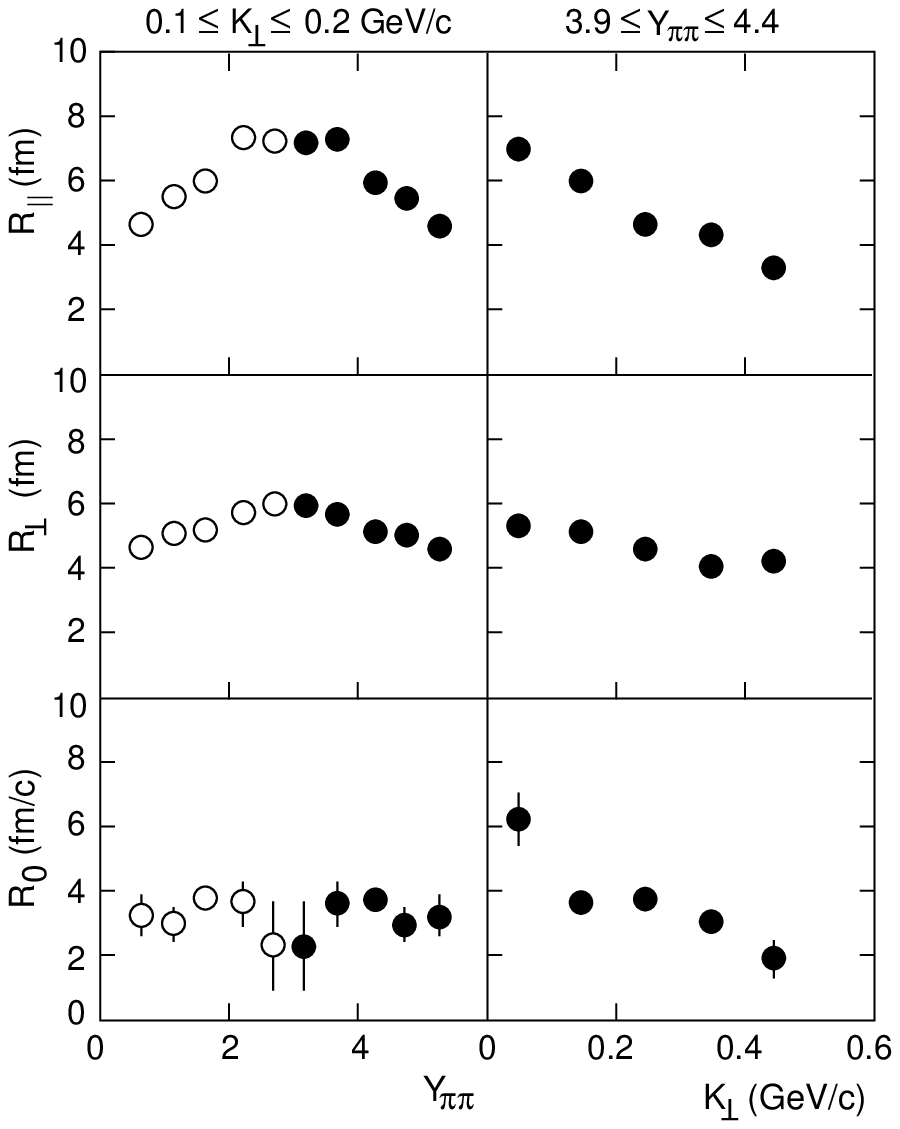}
	 \hfill
   \end{minipage}
   \hfill  
   \vspace*{-1.0cm}
   \begin{minipage}[b]{6.0truecm}
	 \epsfxsize 6.0truecm \epsfbox{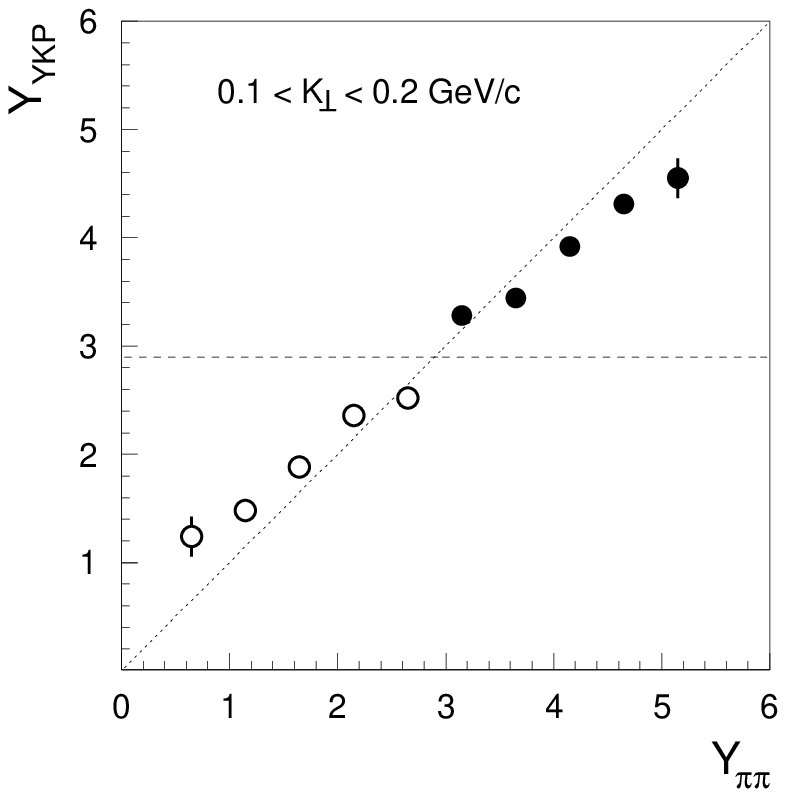}
	 \hfill
   \end{minipage}
\end{center}
\caption{Left: YKP radius parameters for pion pairs from 158 $A$ GeV 
   Pb+Pb collisions measured by NA49, as functions of the transverse 
   momentum $K_\perp$ and rapidity $Y\equiv Y_{\pi\pi}$ as indicated 
   in the Figure. Right: Effective source rapidity $Y_{_{\rm YK}}$ for 
   pions as a function of pair rapidity $Y\equiv Y_{\pi\pi}$, for pairs
   with small transverse momenta ($0.1<K_\perp<0.2$ GeV). The dashed 
   horizontal line indicates the expectation from a non-expanding source.
   Filled circles are measured data, open circles are reflected about 
   midrapidity ($Y_{\pi\pi}=2.9$). (Figure taken with permission 
   from \cite{NA49HBT}.)  
\label{NA49YKP}}
\end{figure}

The left part of Figure~\ref{NA49YKP} shows the fit results for the 
parameters $R_\parallel, R_\perp$ and $R_0$, both as a function of pair 
rapidity $Y\equiv Y_{\pi\pi}$ for small transverse pair momentum 
$0.1 < K_\perp < 0.2$ GeV/$c$ and as a function of $K_\perp$ for 
forward moving pairs at $3.9 < Y < 4.4$. 
%
%
$R_\parallel$ and $R_\perp$ peak at midrapidity, the maximum of 
$R_\parallel$ ($\approx 7-8$ fm) being somewhat larger than that of 
$R_\perp$ ($\approx 6$ fm, more than twice as large as expected from 
the initial nuclear overlap region). Since at midrapidity $R_l=R_\parallel$, 
the NA49 YKP radius parameters $R_\parallel,R_\perp$ can be compared 
with $R_l,R_s$ from NA44; they are consistent. While all three YKP radius 
parameters are decreasing functions of $K_\perp$, the $K_\perp$-dependence 
of $R_\parallel$ is clearly stronger than that of $R_\perp$, indicating 
dominant longitudinal and somewhat weaker transverse expansion. 

The right diagram in Figure~\ref{NA49YKP} shows a strong correlation
between the effective source rapidity $Y_{_{\rm YK}}$ extracted from 
the YKP fit and the pair rapidity $Y_{\pi\pi}$. Following the 
discussion in Sec.~\ref{sec2.3.1} this is again evidence for very 
strong longitudinal expansion of the source. The data points in the 
Figure seem to indicate a slope slightly below unity, as expected from 
thermal smearing effects at low $K_\perp$ (see Figure~\ref{F2}).
 
The data were compared \cite{NA49HBT} to the expanding source 
model (\ref{3.15}). In each $Y$-bin the $K_\perp$-dependence of the 
$R$-parameters can be successfully described by the model, and 
in particular the strong $K_\perp$-dependence is reproduced very well 
by the assumption of {\em boost-invariant} longitudinal flow.
The rapidity dependence of $R_\perp$, however, cannot be reproduced 
by a constant transverse Gaussian radius $R$ of the source; it requires 
$R$ (and thereby the average transverse flow) to decrease away from 
midrapidity \cite{CN96,D99}. Near midrapidity one finds for the model 
parameter $R$ in (\ref{3.15}) $R\approx 8$ fm, shrinking to $R \approx 7$ 
fm at $3.9<Y<4.4$ \cite{NA49HBT}. From (\ref{Rstar}) one obtains the 
ratio $\eta_f^2/T = 3.7 \pm 1.6$ GeV$^{-1}$ \cite{NA49HBT}.
For freeze-out temperatures in the range 100-140 MeV this implies 
transverse flow rapidities $\eta_f$ of 0.6-0.72, corresponding to 
average transverse flow velocities of 0.5-0.6$c$.


\subsubsection{Emission duration}
\label{sec4.5.3}

In sulphur-induced collisions at the SPS \cite{na35zeit,na44mt} and 
in Pb+Pb collisions analyzed by NA44 with the Cartesian parameterization 
\cite{na44pb} the emission duration was found to be very short --
consistent with 0-2 fm/$c$ (with considerable statistical and systematic 
uncertainties). NA49 found from their YKP fit to the Pb+Pb data 
a non-zero emission duration of approximately 3 fm/$c$ \cite{NA49HBT}. 
All these numbers are rather short compared to the emission duration 
predicted by RQMD, namely 3-6 fm/$c$ for S+Pb \cite{fields} and 7-8.5 
fm/$c$ for Pb+Pb \cite{na44pb}, despite the fact that RQMD provides a 
good representation of the particle distributions. This illustrates the 
difficulty and model dependence of extracting the emission duration in 
relativistic heavy ion collisions, discussed in the previous sections.

This is different for low-energy heavy-ion collisions at $E/A$=30-80 MeV 
where emission durations of up to 1400 fm/$c$ were measured \cite{L93} -- the 
typical evaporation time of a compound nucleus. Such large times can be 
extracted with relatively much less model uncertainty. If the creation of
a quark-gluon plasma led to a very long-lived intermediate stage near the
critical temperature $T_c$ for hadronization, it could emit hadrons from 
the surface over much longer periods of time than presently measured.
This might leave more easily interpretable traces in $R_{\rm diff}^2$ or 
$R_0^2$ \cite{BGT88,B89,RG96}.


\section{COMBINING SINGLE- AND TWO-PARTICLE SPECTRA}
\label{sec5}

It has been shown that the single-particle $m_\perp$-distributions of 
pions and heavier hadrons reflect a transversely expanding source 
\cite{NA44flow,pbmswx,AGSspectra}. Such spectra are sensitive to a 
different combination of $T$ and $\eta_f$ \cite{LH89,NA44flow}.
Combining this information with that from the two-particle 
correlation functions thus allows a separation of the collective 
and thermal momentum components described by $\eta_f$ and $T$
\cite{CLZ94,CL95,bjQM95,CN96}.

%
\begin{figure}[ht]
\vspace*{7cm}
\includegraphics{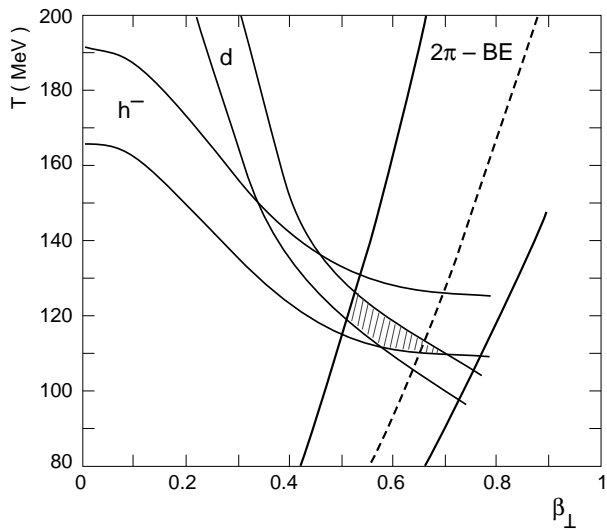}
\caption{Allowed regions of freeze-out temperature $T$ and transverse
         flow rapidity $\eta_f$ (here denoted as $\beta_\perp$) for central
         Pb+Pb collisions near midrapidity. The bands labelled $h^-$ and
         $d$ indicate fits to the negative hadron and deuteron spectra; 
         the third band stems from a fit to the sideward radius parameter
         from 2-pion correlations. (Figure taken with permission
         from \protect\cite{NA49HBT}.)
\label{TvsB}}
\end{figure} 
%

NA49 has analyzed negative hadron and deuteron spectra in Pb+Pb collisions
at the SPS and combined the fit parameters with their correlation function 
analysis near midrapidity \cite{NA49HBT}. Figure~\ref{TvsB} shows the 
allowed regions of freeze-out temperature and radial velocity as bands 
of $\pm \sigma$ around each of the three independent fits. These favor 
a narrowly defined overlap region with $T = 120 \pm 12$ MeV 
and $\eta_f = 0.55 \pm 0.12$ \cite{NA49HBT}. From the model parameter,
$\eta_f$, one can calculate the average transverse flow velocity 
$\langle v_\perp \rangle$ (see below), given the density and velocity 
profiles. The fit to the $h^-$ 
spectrum in \cite{NA49HBT} was made without explicit consideration 
of the contributions from resonance decays and heavier hadrons; once 
these are included \cite{WTH98}, the band labelled ``$h^-$'' in 
Figure~\ref{TvsB} no longer bends over at small $T$, and the crossing 
region is shifted slightly downward.

A {\em simultaneous} fit to all the available one- and two-particle
distributions was done on Si+Au data from the AGS by Chapman and 
Nix \cite{CN96}. They found a freeze-out temperature $T\simeq 90{-}95$ 
MeV and an average radial flow velocity of $\langle v_\perp \rangle 
\simeq 0.34\,c$. A comprehensive analysis of single-particle spectra from 
Au+Au collisions at the AGS (for which the 2-particle correlations 
still await publication) confirms \cite{D99} the low freeze-out 
temperature $T\approx 90$ MeV, with an even larger average transverse 
flow $\langle v_\perp \rangle \approx 0.45\,c$. A preliminary simultaneous
analysis of spectra and correlations from Pb+Pb collisions at the SPS 
measured by NA44 \cite{HvH} gave $T\sim 95{-}100$ MeV and $\langle v_\perp 
\rangle \approx 0.34\,c$. A simultaneous fit to the NA49 Pb+Pb $h^-$ 
spectra and correlations \cite{WTH99} yielded $T\approx 100$ MeV with 
$\eta_f=0.6$, corresponding to an average transverse expansion 
velocity $\langle v_\perp \rangle \approx 0.5\,c$. These freeze-out 
temperatures are somewhat lower (and the radial expansion velocity 
correspondingly higher) than the values extracted from single particle 
spectra alone \cite{pbmswx,NA44flow,K97}. 

Knowing the average transverse flow velocity $\langle v_\perp \rangle$ 
and the transverse size $R$ at freeze-out, a lower estimate for the total 
expansion time can be obtained. Comparing for the Pb+Pb collisions at the
SPS the 1-d rms radius at midrapidity, $R\approx 8$ fm (see 
Sec.~\ref{sec4.5.2}), with the corresponding value of a cold Pb 
nucleus, 3.2 fm, one arrives with $\langle v_\perp \rangle \lapp 0.5\,c$ 
at a lower limit of 10 fm/$c$ for the total duration of the transverse 
expansion. This is larger than the value $\tau_0\approx 8$ fm/$c$
extracted \cite{NA49HBT} from the measured midrapidity value of 
$R_\parallel$ using (\ref{LH}). This illustrates the already mentioned 
problems of interpreting $\tau_0$ directly as the total expansion time, 
and it also indicates that the longitudinal expansion was not always 
boost-invariant, but that the system underwent longitudinal 
acceleration before freeze-out.


\section{GLOBAL TRENDS}
\label{sec6}
\subsection{From p+p to Pb+Pb Collisions}
\label{sec6.1}

Analysis of correlation functions in p+p and $\pi/K$+p collisions yields 
source parameters $R$ of approximately 1 fm \cite{pp,pip}. Proton 
collisions upon nuclei yield larger effective sources. As shown above, 
$R_s$ from p+Pb collisions at 450 GeV is approximately 2 fm, 
smaller than the corresponding value (3.2 fm) for a cold Pb nucleus. 
The density of produced particles in such collisions is not large and 
so significant expansion should not be expected; the increase relative 
to pp is more likely due to cascading of the struck target nucleons.

In heavy-ion collisions at both AGS and SPS energies significant 
flow is observed via the $M_\perp$ dependence of the source parameters.
Consequently, $R_s$ measured with pairs of finite $M_\perp$ provides 
only a lower limit to the actual size of the source. Even so, the 
measured values reflect {\em effective} sources which are already 
considerably larger than the initial nuclear overlap region. 
For Si+Au collisions at 14.6 GeV/nucleon, the ratio $\xi$ of $R_s$ 
to $R^{\rm cold}_s$ (as defined after Eq.~(\ref{eq3.2})) shows that 
the source expands by at least a factor $\xi=1.8$. As 
their measured correlation functions agreed with those from the 
RMQD model, the E814 collaboration used RQMD to correct for the 
flow effects \cite{814l}. They inferred an expansion factor 
$\xi= 2.7$ by comparing $R_t = \sqrt{\langle x^2 + y^2\rangle}$ 
from RQMD with the 2-dimensional radius of the projectile in the 
transverse plane (see Sec.~\ref{sec4.1.5}).

For Au+Au collisions at the AGS, the lower limit on the expansion 
factor from $R_s$ is $\xi\gapp 1.6$, while for Pb+Pb at CERN it is 
$\xi\approx 2$. Data from symmetric collisions of Fe, Nb, and La at 
the Bevalac \cite{bossy,chacon} at 1.3-1.7 GeV/nucleon indicate 
expansion factors $\xi$ of at least 1.2--1.4.

\subsection{Beam Energy Dependence}
\label{sec6.2}

Central Au+Au collisions at 2, 4, 6, and 8 GeV/A were measured by E895
\cite{MLisa} and analyzed with the parametrizations described in 
Sec.~\ref{sec2.2.2}. Though the analysis is still preliminary, some 
very interesting trends are evident. The $K_\perp$-dependence of $R_s$ 
characteristic of radial flow is observed in all but the lowest 
energy collisions, where pion statistics limit the measurement. 
Longitudinal flow (via $R_\parallel(K_\perp)$) is seen at all 
energies. The transverse expansion factor $\xi$ ranges from 1.5 to 2.0. 
It is intriguing that the {\em largest} expansion is at the 
{\em lowest} energy, where the relative importance of the radial
flow appears to be smallest. It is tempting to conclude that lower 
expansion velocity coupled with large final size implies a 
long-lived source. However, the data indicate that the emission 
duration in all cases is quite short, leaving one to wonder why 
surface emission of pions appears to be missing.

Direct comparisons with Bevalac data are complicated by the inability 
to select the most central collisions. However, the fits use the function
$C(q_\perp,q_\parallel,q^0) = 1 + \lambda \exp[(-q_\perp^2 R_\perp^2 
- q_\parallel^2 R_\parallel^2 - (q^0)^2 \tau^2)/2]$, so $R_\perp$ may 
be compared with $R_s$. Studies of 1.7 GeV/A Fe+Fe, 
1.5 GeV/A Nb+Nb \cite{chacon} and 1.3 GeV/A La+La collisions \cite{bossy} 
yield $R_\perp$ values (corrected by $\sqrt{2}$ for comparison 
with the AGS fits) of 2.8, 3.4, and 3.2 fm, respectively. These 
give $\xi$ = 1.2--1.4. The disagreement with the low-energy AGS 
results may be due to the centrality difference of the collisions.

Collecting all of the data together, one may look for trends with 
$\sqrt{s}$ in the region between 2 and 20 GeV covered by present 
data. For asymmetric heavy ion collisions with small projectiles, 
$R_s$ or $R_\perp$ is always 3-4 fm, regardless of $\sqrt{s}$. For 
small symmetric systems ($A<100$), $R_s$ is likewise approximately 
independent of $\sqrt{s}$, in the range 2.5-3.5 fm. For symmetric 
systems with $A>100$, it is more instructive to look at the expansion 
factor $\xi$ which appears to increase with $\sqrt{s}$. This conclusion 
relies heavily, however, upon the La+La measurement \cite{bossy} which 
is low ($\xi=1.2$).

\section{CONCLUSIONS AND FUTURE PERSPECTIVES}
\label{sec7}
\subsection{Where Do We Stand?}
\label{sec7.1}

In this review we have described the development of a sophisticated 
framework with which to extract physics from two-particle correlation 
measurements in heavy-ion collisions. Theoretical and experimental 
progress in the past decade allows characterization of the particle 
source in these rather complex systems and gives access, for the first 
time, to the dynamical evolution. At last the promise of elucidating
the space-time evolution of the particle source directly from measured
quantities has been realized. These new techniques are now also being 
applied to $e^+e^-$ collisions.

Two-particle correlations measure collective flow of the matter, via the 
pair momentum dependence of the homogeneity region, thus fixing the ratio
of the freezeout temperature and average flow velocities. Combining this 
information with an analysis of the single particle spectra uniquely 
separates temperature and flow. The detailed characterization of the 
final state which is now possible provides stringent constraints on 
models simulating the dynamical evolution of the reaction zone.

\subsubsection{Heavy-ion collision dynamics}
\label{sec7.1.1}

Analysis of correlation functions has shown that tremendous expansion
of the system takes place before the hadrons decouple. Longitudinally
the source at freeze-out features approximately boost-invariant flow
while the transverse dynamics is slightly weaker. Still, the transverse 
radius of the particle emitting source approximately doubles from initial 
impact to freeze-out. These features indicate action of a significant 
pressure, though the hadronic observables are not able to indicate which 
degrees of freedom are responsible for its build-up. 

Already at Bevalac energies, below 2 GeV/nucleon, the hadronic matter 
expands. However, both the radial flow velocity of the matter and the 
final source size tend to increase with increasing beam energy. High 
energy collisions at the SPS result in hadron sources which develop 
average transverse flow velocities of 0.5\,$c$; the hadrons freeze out 
at temperatures near 100 MeV. Kaons and pions flow together, and the 
observations are consistent with freeze-out from a common source. From 
the onset of expansion to freeze-out, the transverse radius increases 
by a factor of 2.5; given the difference in expansion velocities, the 
longitudinal growth should be about twice that. The total source 
volume in Pb+Pb collisions thus grows by a factor of 30!

\subsubsection{Initial conditions}
\label{sec7.1.2}

The single and two-particle distributions are consistent with 
formation of a thermalized, flowing hadron gas. The behavior of such 
a gas may be used to extrapolate back to the early times in the 
collision, using the {\em measured} freeze-out conditions and flow 
gradients constrained by the data. The freeze-out temperature, via the
equation of state of an ideal resonance gas, provides an estimate
for the local energy density at decoupling ($\approx 80$ MeV/fm$^3$ at
$T\approx 100$ MeV), the measured transverse flow velocity provides the
Lorentz contraction factor $\gamma^2$ to correct for the kinetic energy
of the expanding matter in the lab frame ($\gamma^2\approx 1.3$ for
$\langle v_\perp\rangle \approx 0.5$). This gives a freeze-out energy 
density of about 100 MeV/fm$^3$. With the total expansion factor 30
given above, the initial energy density, at the onset of transverse 
expansion, should have been of the order of 3 GeV/fm$^3$, {\it i.e.} 
well above the critical energy density $\epsilon_s\lapp 1$ GeV/fm$^3$ 
for color deconfinement as given by lattice calculations \cite{Karsch}. 
Since the initialization of 
transverse expansion requires pressure, this large energy density must 
have been at least partially thermalized. This estimate of the initial
energy density (which is averaged over the transverse area of the source)
agrees in order of magnitude with simple estimates which use the
Bjorken formula \cite{bj} with the measured multiplicity density and
assume an initial thermalization time $\tau_{\rm eq} = 1$ fm/$c$. Compared 
to the latter it has, however, the advantage that it replaces several 
features of the highly idealized Bjorken expansion model \cite{bj}
(e.g. the parameter $\tau_{\rm eq}$) by measured quantities, extracted 
from the HBT analysis.  

\subsection{The Future}
\label{sec7.2}

Experimental measurements in the coming years will develop in several 
directions. Let us mention a few in which serious activities are 
already now visible:

\subsubsection{Heavy-ion collisions at higher energies}
\label{sec7.2.1}

Higher energy collisions (factor of 10 increase in $\sqrt{s}$) will be 
available at the Relativistic Heavy Ion Collider (RHIC) in late 1999. 
Experiments will use the techniques described here to map the 
freeze-out conditions. Quantifying the transverse and longitudinal 
flow velocities will allow determination of the pressure build-up in 
the early stages of the collisions. Armed with the extracted velocities, 
one will be able to back-extrapolate from freeze-out to the time of 
hadron formation. Full analysis of kaon and proton correlations both 
extends the accessible $K_\perp$ range and will verify whether these 
hadrons are emitted from a common source with the pions. 

\subsubsection{Azimuthally sensitive HBT analysis}
\label{sec7.2.2}

We have only discussed the analysis of azimuthally symmetric 
sources. Even if impact parameter $\bbox{b}=0$ never occurs,
the ``central'' event ensemble 
from which the correlation function is constructed is azimuthally 
symmetric since one averages over the orientation of the collision 
plane. On the other hand, reconstructing the latter event-by-event
opens up an even richer field of activities and phenomena. Azimuthal 
anisotropies of global event features (in particular directed and 
elliptic flow, for reviews see \cite{SG86,Olli98}) have been studied 
since the days of the Bevalac, and azimuthally sensitive analyses of 
single-particle spectra from non-central collisions have recently 
attracted a lot of attention \cite{E877aniso}. The measured elliptic
flow of pions and protons at the SPS \cite{NA49flow} may play an 
important role in extracting the pressure in the early collision 
stage \cite{Sorge98}. The next question is how 
these azimuthal deformations in momentum-space are correlated with
corresponding deformations in coordinate-space. The tools for extracting
this information from azimuthally sensitive HBT analyses were developed 
in \cite{VC96,W98a,Heisel98}.  

Without azimuthal symmetry of the source, the correlator in
Gaussian approximation is characterized by 6 functions of 3 variables,
see (\ref{CartHBT}). Following the pioneering studies of Voloshin and
Cleland \cite{VC96}, it was shown by Wiedemann \cite{W98a} that for
sources whose transverse geometric and dynamical deformations have a
dominant quadrupole component this can be reduced to {\em 6 functions  
of only 2 variables} $(K_\perp,Y)$. The dependence on the angle 
$\Phi$ between $\bbox{K}_\perp$ and the impact parameter $\bbox{b}$ 
is made explicit in the {\bf Wie\-de\-mann parametrization}: 
 \begin{eqnarray}
 \label{Wied}
   C(\bbox{q},\bbox{K}) &=& 1 \pm \exp[-R_s^2 q_s^2 -R_o^2 q_o^2
   -R_l^2 q_l^2 - 2 R_{ol}^2 q_o q_l]
 \nonumber\\
   &&\quad\times
   \exp[-\alpha_1\cos\Phi(3q_o^2+q_s^2)+2\alpha_1\sin\Phi q_o q_s]
 \nonumber\\
   &&\quad\times
   \exp[-\alpha_2\cos(2\Phi)(q_o^2-q_s^2) 
        + 2\alpha_2\sin(2\Phi)q_o q_s]\, .
 \end{eqnarray}
The parameters $R_s,R_o,R_l,R_{ol}$ are the same as in (\ref{PBC});
they are given by the $\Phi$-averages of the corresponding functions
of $\bbox{K}{=}(Y,K_\perp,\Phi)$ in (\ref{Cartesian}) and describe the
homogeneity regions of the azimuthally averaged deformed source. The
two new parameters $\alpha_{1,2}(K_\perp,Y)$ are related to the first
and second harmonic coefficients of $R_s$ in the azimuthal angle
$\Phi$. They were shown \cite{W98a} to characterize dynamic and
geometric elliptic anisotropies in the source, respectively. 

\subsubsection{The average phase-space density at freeze-out}
\label{sec7.2.3}

Determination of the phase-space density of particles at freeze-out 
will be important to check whether the collisions at higher energies 
produce a source consistent with local thermal equilibrium. It has been
speculated that the higher densities of produced particles might cause a
pion condensate \cite{P93}. Here, a comparison of pion and kaon 
phase-space densities may be particularly illuminating.

The average phase-space density of pions at freeze-out has already 
been studied at AGS and SPS energies and found to be consistent with
particle emission from a source in local thermal equilibrium.
Using Eq.~(\ref{psdens}), E877 analyzed positive and negative pions 
from 10.8 $A$ GeV Au+Au collisions near beam rapidity. They 
found~\cite{M96} that transverse flow is not needed, and the data are 
well-described by a thermal source. Study of a compilation \cite{FTH98} 
of data from S+nucleus and Pb+Pb collisions at the SPS showed a nearly 
universal behavior of the phase-space density of pions at freeze-out. The 
data are again consistent with freeze-out from a locally equilibrated 
source. However, the $p_\perp$-dependence of the spatially averaged 
phase-space density at mid-rapidity is less steep than that of a 
Bose-Einstein distribution. Such a deviation is expected due to 
transverse flow.

\subsubsection{Three-pion correlations}
\label{sec7.2.4}

Higher order Bose-Einstein correlations will be studied to seek evidence 
of cohe\-rence in the source. An analysis as described in Sec.~\ref{sec2.6} 
was done by NA44~\cite{na443pi}, albeit due to limited statistics only in 
one dimension. A three-pion correlation 
function $C_3(q_3) = 1 + \lambda_3 \exp(-R_3^2 Q_3^2)$, 
with $Q_3^2{=}(p_1{-}p_2)^2{+}(p_2{-}p_3)^2{+}(p_3{-}p_1)^2$, 
%
\begin{figure}[ht]
\centerline{\epsfxsize=13cm\epsffile{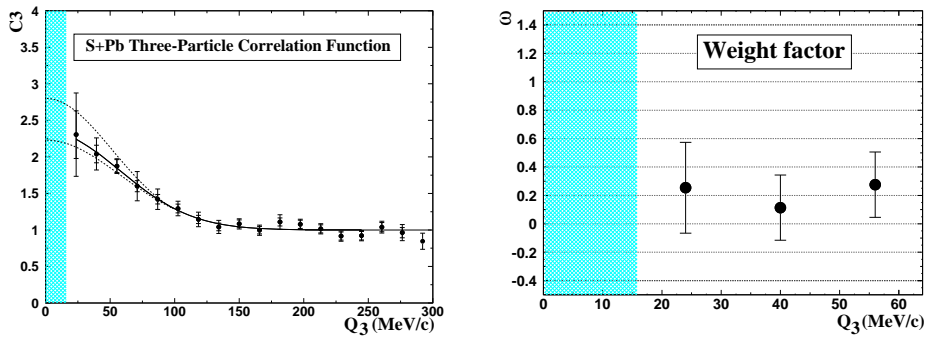}}
\caption{Left panel: Three-pion correlation function from S+Pb 
         collisions measured by NA44, as a function
         of $Q_3$ (see text). Right panel: Half the normalized true 3-pion
         correlator $r_3$ (here called $\omega$) as a function of $Q_3$.
         (Figure taken with permission from \protect\cite{na443pi}.)
\label{3pi}}
\end{figure} 
%
was constructed from central S+Pb collisions at 200 $A$ GeV; it is 
shown on the left side of Figure~\ref{3pi}. The upper dashed line 
shows the three-pion correlation function expected if the source is
totally chaotic and symmetric, while the lower shows the case of 
vanishing $r_3$ (see Sec.~\ref{sec2.6}). The right side of the 
Figure shows that the normalized three-pion correlation function
$r_3$ (which is experimentally determined as a weight factor called 
$\omega$ \cite{na443pi}) deviates from 1 in the experiment in the 
accessible $Q_3$-range; if this behaviour persisted down to $Q_3$=0, 
this would imply partial coherence of the source. It is hoped that 
the much higher statistics expected at RHIC will settle this 
important question.

\subsubsection*{\bf Acknowledgments}

We would like to thank the Institute for Nuclear Theory at the 
University of Washington, where this work was started, for its 
hospitality and the stimulating environment it provided. We
appreciate help from M. Kopytine in the
preparation of figures. We gratefully acknowledge T. Cs\"org\H o,
H. Heiselberg, M. Lisa and U.A. Wiedemann for careful reading
of the manuscript and constructive comments.
This work was supported in part by DFG, 
BMBF, GSI and the U.S. Department of Energy, under Grant 
DE-FG02-96ER40988.A001.

\bigskip
%

\end{document}